\documentclass[twocolumn]{aastex631}

\newcommand\virga{\texttt{Virga}}

\usepackage{booktabs}
\usepackage{graphicx}
\usepackage[version=4]{mhchem}

\newcommand\picaso{\texttt{PICASO}}
\newcommand\sonorabob{\texttt{Sonora Bobcat}}
\newcommand\cristo{$\beta$-cristobalite}
\newcommand\tridy{$\beta$-tridymite}
\defcitealias{Grant2023clouds}{G23}
\defcitealias{zeidler}{Zeidler et al.}
\defcitealias{Henning1997}{Henning \& Mutschke}
\usepackage{tabulary}
\newcommand\chisq{$\chi^2_{\nu}$}
\newcommand\fsed{$f_{\rm{sed}}$}
\newcommand\kzz{$K_{\rm{zz}}$}

\revised{August 9, 2024}

\submitjournal{ApJ Letters}

\shorttitle{Substellar Silica Cloud Polymorphs}
\shortauthors{Moran, Marley, \& Crossley}

\begin{document}

\title{Neglected Silicon Dioxide Polymorphs as Clouds in Substellar Atmospheres}

\author[0000-0002-6721-3284]{Sarah E. Moran}
\affiliation{Department of Planetary Sciences and Lunar and Planetary Laboratory, University of Arizona, Tuscon, AZ, USA}

\author[0000-0002-5251-2943]{Mark S. Marley}
\affiliation{Department of Planetary Sciences and Lunar and Planetary Laboratory, University of Arizona, Tuscon, AZ, USA}

\author[0000-0002-0619-0197]{Samuel D. Crossley}
\affiliation{Department of Planetary Sciences and Lunar and Planetary Laboratory, University of Arizona, Tuscon, AZ, USA}



\begin{abstract}

Direct mid-infrared signatures of silicate clouds in substellar atmospheres were first detected in {\em Spitzer} observations of brown dwarfs, although their existence was previously inferred from near-infrared spectra. With JWST's Mid-Infrared Instrument (MIRI) instrument, we can now more deeply probe silicate features from 8 to 10 $\mu$m, exploring specific particle composition, size, and structure. Recent characterization efforts have led to the identification of silica (silicon dioxide, SiO$_2$) cloud features in brown dwarfs and giant exoplanets. Previous modeling, motivated by chemical equilibrium, has primarily focused on magnesium silicates (forsterite, enstatite), crystalline quartz, and amorphous silica to match observations. Here, we explore the previously neglected possibility that other crystalline structures of silica, i.e. polymorphs, may be more likely to form at the pressure and temperature conditions of substellar upper atmospheres. We evaluate JWST’s diagnostic potential for these polymorphs and find that existing published transmission data are only able to conclusively distinguish tridymite, but future higher signal-to-noise transmission observations, directly imaged planet observations, and brown dwarf observations may be able to disentangle all four of the silica polymorphs. We ultimately propose that accounting for the distinct opacities arising from the possible crystalline structure of cloud materials may act as a powerful, observable diagnostic tracer of atmospheric conditions, where particle crystallinity records the history of the atmospheric regions through which clouds formed and evolved. Finally, we highlight that high fidelity, accurate laboratory measurements of silica polymorphs are critically needed to draw meaningful conclusions about the identities and structures of clouds in substellar atmospheres.

\end{abstract}

\section{Introduction} 



Substellar objects are fundamentally about clouds. Clouds were understood as likely being present almost from the beginning of brown dwarf studies and are part of the reason that Jill Tarter even created the ``brown" moniker, even though ``brown is not a color" \citep{Tarter86}. Silicates and other astrophysical ``dust'' 
 were identified as the likely condensing species in objects from 1000 to 2000 K via thermochemical equilibrium calculations \citep[e.g.,][]{FegleyLodders96}.  Over the history of brown dwarf science -- and over that of its successor and cousin hot Jupiter science -- the field has subsequently striven to do a progressively better job modeling such clouds \citep[e.g.,][]{Lunine1986,Sharp1990,Tsuiji1996,ackerman2001,Calamari2024}. 

Observationally, it was clear from optical and near-infared slopes that condensate grains --- clouds --- had to be present in these warm substellar atmospheres \citep[e.g.,][]{Marley2002,Cushing2006}. However, only with InfraRed Spectrograph (IRS) Spitzer observations did the distinctive silicate feature around 8 -- 10 $\mu$m emerge observationally as a clear indicator of cloud composition \citep{Roellig2004,Cushing2006,Looper2008,Stephens2009,Suarez2022} in L dwarf atmospheres.  

This silicate feature arises from the vibrational mode of the diatomic Si--O bond
that is common to all silicates, which includes magnesium silicates (\ce{MgSiO3}, \ce{Mg2SiO4}), pure silicates (SiO, \ce{SiO2}), calcium- (\ce{Ca2Al2SiO7}, \ce{Ca2SiO4}, \ce{CaMgSi2O6}, \ce{Ca2MgSi2O7}) and iron-bearing silicates (\ce{Fe2SiO4}, \ce{FeSiO3}), and intermediary cases from the solid solution such as olivine and pyroxene \citep{BurrowsSharp99,WakefordSing2015,Luna&Morley2021}. The feature is also commonly observed in cometary comae \citep{Hanner1994}. 

Silicate clouds were also quickly recognized as likely in hot Jupiter atmospheres from theory, Spitzer, and Hubble Space Telescope (HST) observations \citep[e.g.,][]{Seager2000,Richardson2007, WakefordSing2015,Lee2016,Gao2020}. However, the loss of Spitzer's longer wavelength mode in the Warm Spitzer era meant that observational confirmation for such cloud compositions awaited JWST's mid-infared capabilities.


From thermochemical equilibrium and nucleation efficiency calculations, magnesium silicates (i.e., \ce{MgSiO3}, enstatite and \ce{Mg2SiO4}, forsterite) were thought to be the likeliest composition of condensed particles near the photosphere of these objects warmer than $\sim$1000 K \citep{FegleyLodders96,BurrowsSharp99,Gao2020}. Significant efforts that followed focused on the particle morphology differences between crystalline and amorphous forms of enstatite and forsterite \citep{Cushing2006,Helling2006}, which was recognized as likely possible to observationally distinguish with JWST \citep{Luna&Morley2021}. The question of particle morphology and whether silicates form and persist as amorphous or crystalline particles is the subject of some tension. Observational efforts have found evidence for amorphous forms \citep[e.g.,][]{Cushing2006,Burningham2021,Dyrek2024} while high temperature, low pressure laboratory studies suggest crystalline formation via annealing should quickly occur even if particles form as amorphous initially \citep{Fabian2000,Jaeger2003,Toppani2006,Koike2013}.

Notably, pure silica -- \ce{SiO2} -- was initially discounted in many substellar atmospheric studies due to these thermochemical equilibrium considerations \citep[e.g.,][]{Sharp1990,FegleyLodders96}. Conversely, models that went beyond equilibrium chemistry, invoking ``dirty grains'' with precursor seed particles \citep{HellingWoitke2006}, predicted that \ce{SiO2} condensates were likely abundant \citep{Helling2006,Lee2016}.

Recently, in both brown dwarfs and hot Jupiters, specific detections have been made that strongly favor pure \ce{SiO2} over magnesium silicate clouds alone. \citet{Burningham2021} performed an extensive atmospheric retrieval study of the L4.5 dwarf 2MASSW J2224438-015852 and found strong evidence for a quartz cloud co-existing with an enstatite cloud above 0.1 bar, both made of sub-micron sized particles. 

Since the ongoing science mission of JWST, several JWST/MIRI programs have seen evidence for silicate clouds generally \citep{Miles2023,Dyrek2024,Welbanks2024} but could not pinpoint the exact composition of the cloud due to the broadness of the absorption feature observed in the 8 -- 10 $\mu$m region. 
Critically, \citet{Grant2023clouds}, hereafter \citetalias{Grant2023clouds}, reported the first JWST/MIRI observations that confidently identified a high altitude pure quartz cloud layer made of nanometer-sized grains, which gives rise to a strong, sharp absorption feature in the transmission spectrum of the hot Jupiter WASP-17 b. 

The identification of quartz in  addition to magnesium silicates has driven new theories about the condensation sequence of refractory materials in substellar objects. In particular, \citet{Burningham2021} noted that the object in which they detected the quartz cloud layer has a Mg/Si ratio of $\sim$0.69, far less than 1, which can readily shift the dominant equilibrium reservoir of silicate into quartz instead of forsterite. \citet{Calamari2024} delved further into the effect of the Mg/Si ratio on cloud composition and oxygen sequestration. They find that the fraction of substellar host stars with a Mg/Si ratio less than 0.9 is quite large, and therefore if companions inherit stellar abundances, quartz clouds should be relatively common in a variety of worlds \citep{Calamari2024}.

\begin{figure*}[ht!]
\centering
{\includegraphics[width=1.0\textwidth]{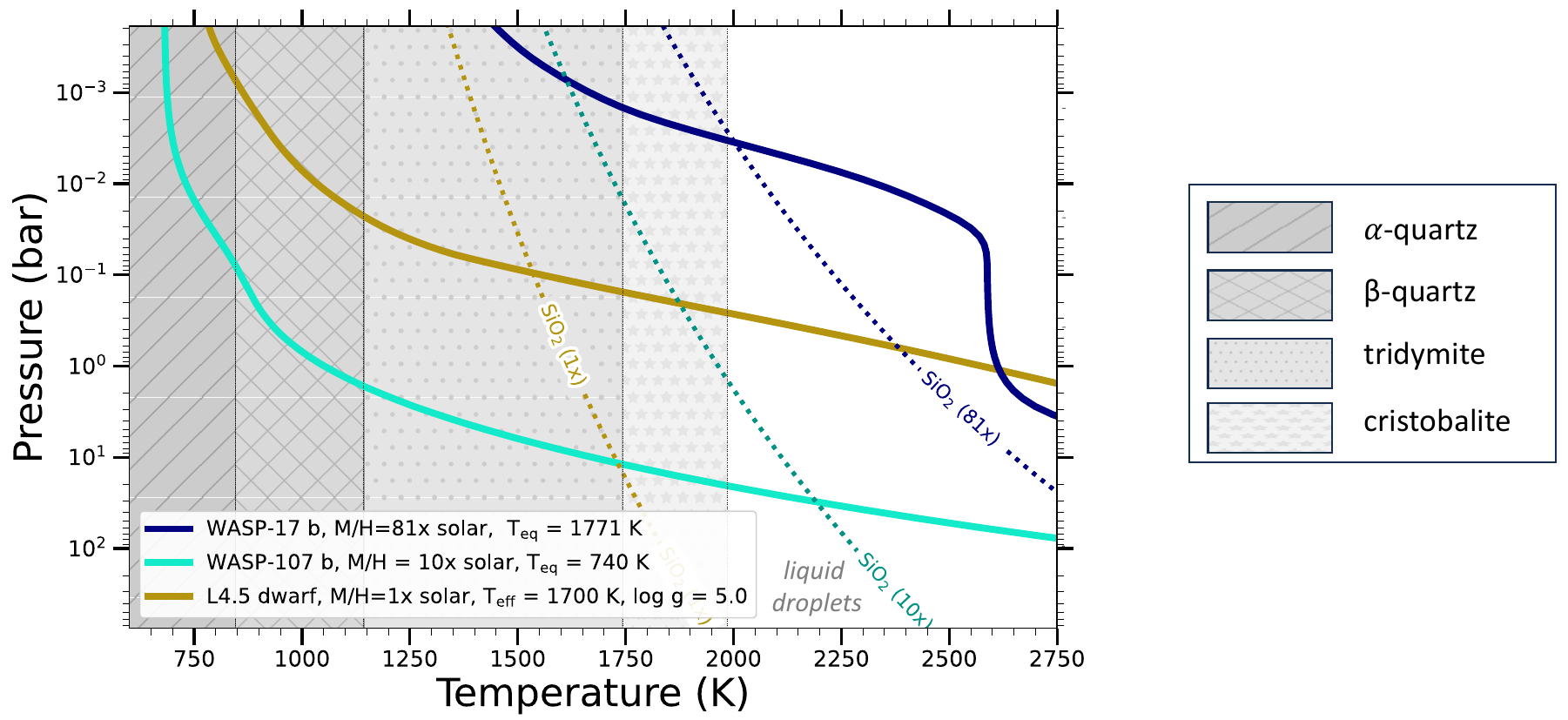}}
\caption{The expected silica polymorphs that form at 1 bar and elevated temperature (indicated by shading according to the legend). Also shown are condensation curves of cloud species for multiple atmospheric metallicities (dashed lines), along with temperature-pressure profiles (solid lines) of the hot Jupiters WASP-107 b (teal; \citealt{Dyrek2024}) and WASP-17 b (blue; \citealt{Grant2023clouds})  and an L4.5 brown dwarf (gold; \citealt{Marley2021}). Where a solid curve crosses a dashed curve of the corresponding color, cloud particles will condense. To the left of a condensation curve but before the cristobalite stability region, liquid cloud droplets should form. \textbf{Quartz, as previously invoked in observational studies, is not the expected first crystalline polymorph of silica to form at warm substellar upper atmospheric conditions.}}
\label{fig:phasediag}
\end{figure*}

Thus, the history of silicate clouds has concentrated strongly on a condensation sequence driven by \textit{chemistry}, with lesser attention given to particle morphologies beyond ``glassy'' or ``crystalline''. Importantly, the literature has neglected significant focus on the \textit{multiple crystalline} structural arrangements possible for silica, each of which has distinct optical properties and thus particle opacity. This neglect comes despite careful consideration given by some authors to accurate optical property computation accounting for all crystallographic axes \citep{KitzmannHeng,Luna&Morley2021}. Indeed, at the high temperatures of brown dwarf and hot Jupiter upper atmospheres, \textit{quartz is not the stable crystalline form of silica}, as shown in Figure \ref{fig:phasediag}.

Instead both $\alpha$- and $\beta$-quartz form below 1143 K at 1 bar, with two alternative crystalline forms that are stable between 1143 K and 1986 K. Note that Figure \ref{fig:phasediag} plots a constant temperature as a function of pressure for each polymorph stability region, as phase data is unavailable below 1 bar, though generally tends asymptotically toward constant at low pressures \citep[e.g.,][]{ManualMineral,Swamy1994,HandbookMinerals,Koike2013}. 

These different crystal arrangements, which are called ``polymorphs,'' have been observed to form in some laboratory experiments examining the formation of silicates in meteorites, cometary grains, young circumstellar environments, and evolved stellar envelopes \citep{Fabian2000,Koike2013}. Consideration of these polymorphs has not yet propagated to the substellar atmospheric literature, which we seek to remedy here.

In this work, we examine two polymorphs of crystalline \ce{SiO2} in addition to amorphous silica and quartz: $\beta$-cristobalite and $\beta$-tridymite. Using the coupled {\picaso} and {\virga} modeling framework, we account for both the optical properties and densities of diverse silicon polymorph cloud particles. In Section \ref{sec:polymorphs} we discuss the physical properties of realistic silica condensates and crystalline structures, in Section \ref{sec:picaso} we discuss our model framework and assumptions, in Section \ref{sec:w17b} we compare our results against the spectrum of WASP-17 b as reported by \citetalias{Grant2023clouds}, and in Section \ref{sec:BD} we predict potential emission spectra with different silica polymorph clouds for a typical L dwarf. Then, in Section \ref{sec:disc_concl}, we explore the limitations of this work and offer suggestions for follow-up efforts considering both experimental and theoretical approaches. We present our conclusions in Section \ref{sec:summary}.

\section{Realistic Silica Polymorphs in Substellar Atmospheres} \label{sec:polymorphs}

\subsection{Substellar Upper Atmosphere Conditions}
When silicate clouds are expected, the observable atmospheres of brown dwarfs and hot Jupiters are typically in excess of 1000 K and, particularly for the latter, below 0.1 bar in pressure. If and when solid or liquid condensates form, they likely settle slowly under gravity to form distinct cloud layers that differ in composition based on the specific temperature conditions of the layer where vapor is transforming to condensate \citep{FegleyLodders96,Visscher2010}. 

Hot Jupiters are strongly radiatively forced by their host stars, which can generate much steeper temperature-pressure profiles compared to L dwarfs, causing multiple condensation curves to come into play for any given atmosphere \citep{Wakeford2017}. Small differences in the temperature-pressure profile can thus shift the dominant species of cloud formation \citep{sing2016}. This sensitivity may also give rise to the possibility not only of distinct cloud layers for any given planet, but also for a distinct dominant crystallinity for any given condensate, much like the diversity of crystal shapes of ice cloud particles in Earth's atmosphere at varying altitudes \citep[e.g.,][]{LibbrectIceCloudReview}.

\subsection{Silica Polymorph Formation and Material Properties}

\begin{table*}[t]
\centering
\caption{Low pressure silica polymorph material properties}
\label{tab:material_props}

\begin{tabular}{@{}l*{5}{l}@{}} 
\toprule 
Name & 
Crystal Habit$^a$ &
Temperature$^a$  &
Density$^b$   &
Refractive Index$^b$  &
Optical Properties \\  
&& (K) & (g cm$^{-3}$) & (at 550 nm) & \\
\midrule
$\alpha$-quartz &
  trigonal &
  $<$ 846 stable &
  2.65 &
  1.55 &
  $<$ 6.25 $\mu$m, \citet{Philipp1985}$^c$ \\ 
  &&&&& $>$ 6.25 $\mu$m, \citet{zeidler}$^d$ \\
  \\
$\beta$-quartz &
  hexagonal &
  846 -- 1143 stable &
  2.53 &
  1.54 &
  $<$ 6.25 $\mu$m, \citet{Philipp1985}$^c$ \\
  &&
  &&& $>$ 6.25 $\mu$m, \citet{zeidler}$^d$ \\
  \\
$\beta$-tridymite &
  hexagonal &
  1143 -- 1743 stable &
  2.22 &
  1.47 &
  $<$ 6.7 $\mu$m, \citet{Philipp1985}$^c$ \\
  && 390 -- 1143 metastable 
  &&& 6.7 -- 9.0 $\mu$m, \citet{lippincott1958infrared} \\
  &&&&& $>$ 9.0 $\mu$m, \citet{Sitzars2000_sio2_ir_spectra}$^e$ \\
  \\
$\beta$-cristobalite &
  cubic &
  $>$ 1743 stable &
  2.20 &
  1.48 &
  $<$ 7.0 $\mu$m, \citet{Philipp1985}$^c$ \\
  && 543 -- 1743 metastable
  &&& $>$ 7.0 $\mu$m, \citet{Koike2013}$^f$ \\
  \\
glassy silica &
  amorphous &
  $\le$ 1300 &
  2.20 &
  1.46 &
  $<$ 6.6 $\mu$m, \citet{Philipp1985} \\
  &&
  &&&$>$ 6.6 $\mu$m, \citet{Henning1997}
  \\ 
\bottomrule
\end{tabular}%
\textit{NOTE--- a)} \citealt{Koike2013} \textit{b)} \citealt{ManualMineral} \textit{c)} Actually $\alpha$-quartz \textit{d)} Actually $\beta$-quartz, measured at 928 K

 \textit{e)} measured at 500 K, amplitude scaled by \citealt{lippincott1958infrared} \textit{f)} annealed at 1773 K, measured at room temperature.

\end{table*}

The various polymorphs of SiO$_2$ are distinguished by their internal crystallographic structures, each of which are stable in distinct pressure-temperature regimes \citep{heaney1994silica}.  At upper atmospheric pressures relevant to gas giant planets ($<$ 1 bar), the stable or metastable crystalline SiO$_2$ polymorphs include quartz, tridymite, and cristobalite (Table \ref{tab:material_props}). Additionally, each polymorph also has low temperature ($\alpha$) and high temperature ($\beta$) structures wherein crystallographic symmetry increases with temperature. As such, the presence of specific SiO$_2$ polymorphs can provide insight into the thermal history of the SiO$_2$ grains. 

For example, SiO$_2$ that crystallized or annealed with a trigonal $\alpha$-quartz structure at low temperature will undergo a phase transition at 846 K to hexagonal $\beta$-quartz accompanied by a volume increase of 0.4\% \citep{Ringdalen2015}. This transition occurs rapidly and reversibly because the relative positions of SiO$_2$ tetrahedra are displaced without breaking atomic bonds. Increasing the temperature further induces a reconstructive phase transition from $\beta$-quartz directly to $\beta$-tridymite at 1143 K with a 14\% increase in volume. Unlike displacive $\alpha$-$\beta$ transitions, reconstructive transitions between different polymorphs break and rearrange atomic bonds in the crystal. Consequently, the reverse transition from metastable tridymite back to quartz is endothermic and kinetically unfavored, requiring slow cooling over long timescales ($\sim10^5$ yrs). This phase transition hysteresis during cooling provides a means of assessing both the peak temperature and cooling history of the grain; identification of high temperature polymorphs indicates that SiO$_2$ grains cooled rapidly from high temperature. The utility of SiO$_2$ polymorphs as recorders of thermal history is exemplified with \textit{in situ} martian sediment analyses, where identification of tridymite provides evidence that SiO$_2$-rich lavas reached the surface, rapidly cooled, and broke down to form sediment at the planet’s surface \citep{Morris2016MarsTridymite,YenTridyMars21, Payre2022MarsTridymiteFormationMech}. 

As phase transitions for SiO$_2$ are primarily temperature dependent, similar inferences about formational temperatures can be made for SiO$_2$ polymorphs in the atmospheres of exoplanets and brown dwarfs. Specific polymorphs may be remotely identified as their distinct crystal geometries are also manifested in their mid-to-far infrared (IR) spectra \citep{Koike2013}. Thus, identification of IR features consistent with a particular polymorph constrains the possible temperature range of its environment to the polymorph's stability field.



In astrophysical contexts, silica formation has been considered primarily in studies of protoplanetary disks \citep[e.g.,][]{Fabian2000,Koike2013,Jang2024ProtoPlaDisk}. Experimental work has shown that nanometer-sized, laser ablated silicate and silica grains (``smokes'') and glasses anneal to high temperature, high symmetry crystalline phases over timescales of minutes to hours to days, with speed of crystallization increasing with temperature \citep{Fabian2000,Jaeger2003}. In protoplanetary disks, the warm inner disk can quickly anneal amorphous silica to crystalline phases before particles experience radial or vertical drift to cooler temperature regions, acting as tracers of astrophysical dust evolution \citep{Jang2024ProtoPlaDisk}. 

These studies assume that silica forms as a shock-quenched amorphous phase, rather than via a high temperature gas phase condensation process, as may be more appropriate for substellar atmospheric cloud formation. Experiments that track this reverse transition, from a metastable high temperature polymorph to a lower but still elevated temperature phase, are lacking. The initial formation and stabilization of crystalline phases happens near instantaneously ($\sim$ 10$^{-12}$ s) if elevated temperatures are maintained \citep{Takada2018}, while glassy or amorphous structures will form via quenching if a droplet is rapidly cooled below its melting temperature (or solidus). Additionally, data is lacking regarding the exact pressure-temperature dependence and stability of these high temperature phases below 1 bar, though consideration has been given to high pressure ($\sim$GPa) regimes for terrestrial exoplanets \citep{duffy2015}.

In substellar atmospheres, the formation of silicate clouds of enstatite, forsterite, and silica are not expected to nucleate directly from vapor to either solid or liquid phases however. Instead, chemical reactions are thought to take place that combine gaseous magnesium, water, and SiO to form enstatite or forsterite \citep{Visscher2010}, while solid silica is thought to nucleate from the chemical reaction of gaseous SiO and oxygen \citep{Grant2023clouds}. Therefore, whether such species actually nucleate -- and as what phase -- depends upon the temperature, pressure, and chemical conditions within a region of atmosphere, which is not fully captured by current laboratory experimental constraints.  If these species are able to form liquid droplets, then the laboratory experiments concerning their quenching behavior to form glassy structures should apply, but whether this occurs is as yet unknown.

\vspace{-6pt}
\subsection{Silica Polymorph Optical Properties}

The bulk of this study relies on the fact that the different crystallinity of polymorphs necessarily means their bond lengths and arrangements differ, which corresponds to differences in spectroscopic absorption and scattering properties. Previous substellar cloud studies have already investigated similar spectral differences between quartz and glassy silica \citep{Cushing2006,Burningham2021,Grant2023clouds}, and we expand upon these studies here.

\citetalias{Grant2023clouds} takes their ``$\alpha$-crystal'' and amorphous silica optical properties for their forward models from \citet{KitzmannHeng}. These values are themselves a compilation of data at 928 K ($\alpha$-crystal, \citealt{zeidler}), 300 K (amorphous silica, \citealt{Henning1997}), and room temperature (both $\alpha$-crystal and amorphous \ce{SiO2}, \citealt{Philipp1985} at wavelengths $<$ 6.25 or 6.6 $\mu$m , not measured by \citetalias{zeidler} and \citetalias{Henning1997}, respectively). For the ``$\alpha$-crystal'', \citet{KitzmannHeng} also account for the anisotropy of the crystal structures, using a mean value for the refractive index across all axes. However, \citet{zeidler} note that their $\alpha$-quartz underwent a phase transition around 850 K to $\beta$-quartz, which is actually more likely at the elevated temperatures of a hot Jupiter upper atmosphere as seen in Figure \ref{fig:phasediag}. Of course, {\tridy} and {\cristo} are even likelier to be stable at high altitudes and temperatures.  

In this work, we continue to use optical properties for amorphous and quartz silica from \citet{KitzmannHeng} as in \citetalias{Grant2023clouds} for simplicity, where ``quartz'' is thus actually some combination of $\alpha$- and $\beta$-quartz, with the $\beta$-phase applicable to wavelengths $>$ 6.25 $\mu$m and the $\alpha$-phase applicable shortward of 6.25 $\mu$m. We encourage future efforts to systematically measure the temperature dependencies and optical properties of silica phases at high resolution across a wide wavelength range.

Extensive wavelength coverage for complex refractive indices of {\cristo} and {\tridy} are lacking, though theoretical high energy ($>$5 eV) refractive indices \citep{Chen2023nkEV} exist, as do numerous experimental infrared spectra. Even then, spectra in the exact range of interest, at relevant temperatures and pressures, are sparse

\subsubsection{Tridymite absorption}
For {\tridy}, absorption spectral data exist both at ambient and elevated temperatures in the mid- and far-infrared ($>$5 $\mu$m) \citep{PlendltridytooIR,EtchepareTooIR78,Cellai95stitch,Sitzars2000_sio2_ir_spectra}. Thermal emission spectra also exist, intended for use in studies of {\tridy} on Mars \citep{Michalski2003MarsTESsilica}. 

However, no wide-wavelength, calibrated infrared spectra at substellar atmospheric temperatures exist, so ultimately, we elect to use {\tridy} data as a combination from the infrared spectra presented in \citet{lippincott1958infrared} and \citet{Sitzars2000_sio2_ir_spectra}. \citet{lippincott1958infrared} provide the most complete data over wavelengths of interest to JWST (2.0 to 15.3 $\mu$m) and is available in physical units. However, these measurements were performed at room temperature, which should correspond to the $\alpha$-phase of tridymite rather than the $\beta$-phase, as would be expected in a high temperature substellar atmosphere.  The data likely also contain contamination due to water vapor in the measurement set-up, observable near 2.9 $\mu$m and 6.5 $\mu$m \citep{lippincott1958infrared}. We therefore exclude values shortward of 6.7 $\mu$m in our analysis. To convert the data to usable values for our {\virga} cloud model, we convert the transmission spectrum from percent transmission $T$ to absorbance $A$ and then to an absorption coefficient $\alpha$ using the Beer Lambert Law:

\begin{equation}
    A = - \log_{10}(T)
\end{equation}

\begin{figure*}[t!]
\centering
{\includegraphics[width=0.99\textwidth]{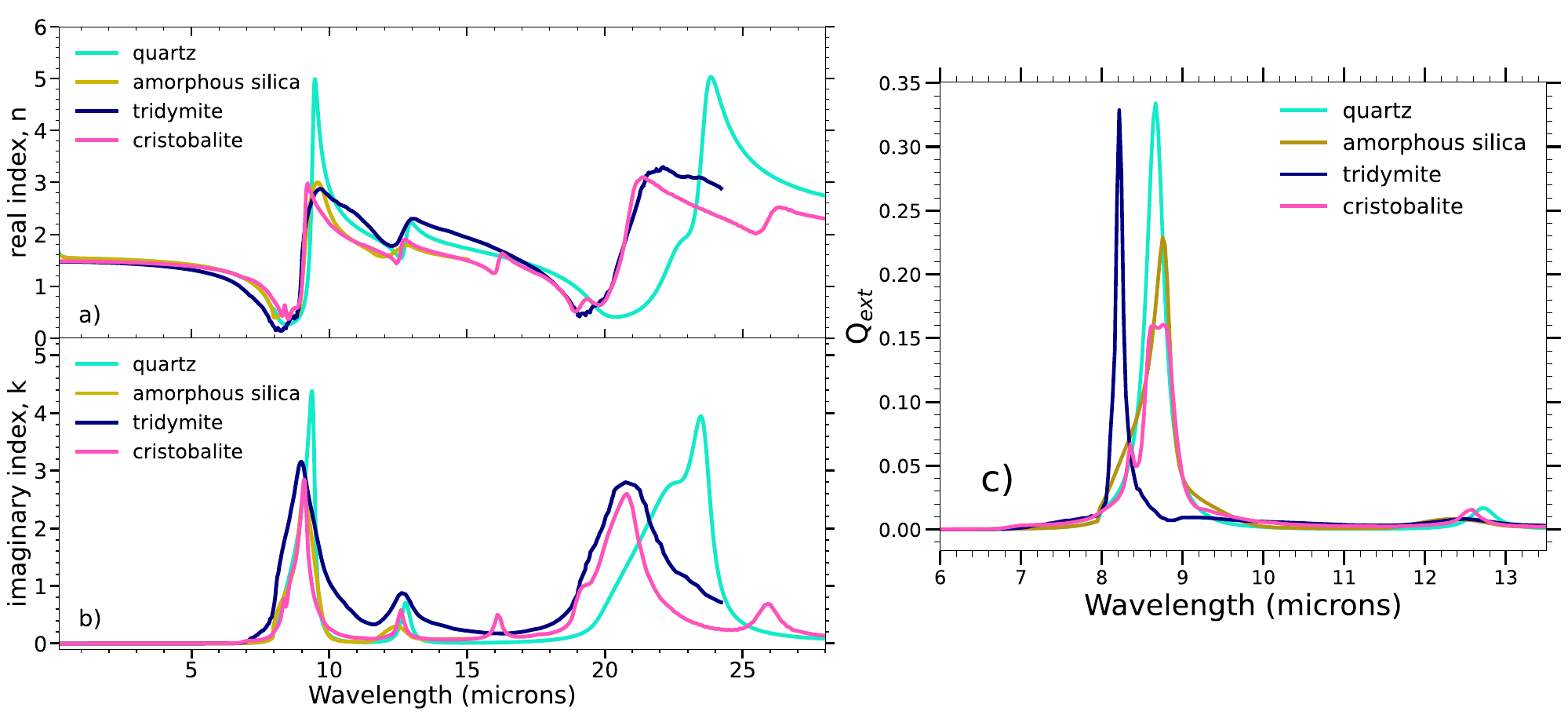}}
\caption{Optical properties of silica polymorphs. a) The real and b) imaginary refractive indices, and c) the extinction efficiencies of various silica polymorphs. Cristobalite and tridymite refractive indices were calculated as described in the text. The extinction efficiencies are shown for a particle radius of 32 nm. \textbf{The different SiO$_2$ crystalline forms have distinct scattering and absorption features that should be separable with the precision of JWST's MIRI instrument.}}
\label{fig:qext}
\end{figure*}

\noindent And then 

\begin{equation}
    \alpha= \ln(10)*A/Ct
\end{equation}

\noindent where $C$ is the sample concentration, given as approximately 0.2\% of tridymite,  and $t$ is the sample thickness, which is not given by \citet{lippincott1958infrared}. However, they used the standard KBr pellet method to obtain their transmission data and we can reasonably estimate 
their pellets were in the range of 0.1 mm thick, with an order of magnitude on the uncertainty in thickness given typical KBr methods. Finally, we can obtain an estimate for the imaginary refractive index, $k$, by the equation

\begin{equation}
   k(\nu) = \frac{\alpha}{4 \pi \nu} \label{imagk}
\end{equation}

\noindent where $\nu$ is wavenumber. \citet{Sitzars2000_sio2_ir_spectra} obtained tridymite spectral measurements from 9 to 24 $\mu$m at high temperatures up to 500 K. We use their spectra for our ultimate calculations, as increasing temperature appears to widen, weaken, and/or shift the peak of silica absorption bands \citep{Cellai95stitch,Sitzars2000_sio2_ir_spectra,zeidler}. However, they do not provide their absorbance data in physical units
. Consequently, we use the $k$ values we computed from \citet{lippincott1958infrared} as a calibration for the amplitude of absorption for the high temperature data, as well as for wavelength values between 6.7 and 9 $\mu$m. Given the sample thickness uncertainty, this propagates out to amplitude uncertainties for k of an order of magnitude as well. However, given the fact that the k amplitudes should not be that different from the other polymorphs, this order of magnitude is a very conservative upper limit. Our calculations using our nominal estimate for the thickness leads to reasonable values as seen in Figure \ref{fig:qext}. A more reasonable uncertainty in k would be on the order of $\sim$20\%.

Since we have no reliable data for {\tridy}'s imaginary refractive index shortward of 6.7 $\mu$m, we simply substitute the imaginary refractive index of $\alpha$-quartz \citep{Philipp1985} for {\tridy} from 0.3 $\mu$m to 6.7 $\mu$m. Should the polymorphs in fact have substantially different imaginary refractive indices at visible and near-infrared wavelengths, our study is thus limited in its application to observations either from JWST's NIRSpec, NIRISS, or NIRCam instruments, or from any observations from HST.

\subsubsection{Cristobalite absorption}
To understand silica in debris disks, \citet{Koike2013} recently measured the mass absorption coefficient, $\kappa$ [cm$^2$ g], for a variety of polymorphs formed at high temperature, including for $\alpha$-cristobalite from 7 $\mu$m to 200 $\mu$m. Therefore, we are able to calculate the absorption coefficient, and thus an estimate of the imaginary refractive index $k$ of {\cristo}. First, we multiply \citet{Koike2013}'s meausured $\kappa$ value for $\alpha$-cristobalite by its material density, 2.33 g cm$^{-3}$, to obtain the absorption coefficient $\alpha$ [cm$^{-1}$] and then apply Equation \ref{imagk} above to obtain $k$. As with {\tridy}, we substitute the imaginary refractive index of $\alpha$-quartz for {\cristo} blueward of 7 $\mu$m where \citet{Koike2013} do not report measurements.

\subsubsection{Tridymite and Cristobalite real refractive indices}
Both {\cristo} and {\tridy} lack measured real refractive indices $n$ across the optical to mid-infrared, we which require in order to calculate Mie coefficients for relevant cloud particle sizes. We therefore perform Kramers-Kronig analysis using the open-source code \texttt{pyElli}\footnote{https://github.com/PyEllips/pyElli} to estimate the real refractive index across the wavelength range of interest. Kramers-Kronig relations take the form

\begin{equation}
    \Delta n(\lambda) = n(\lambda_i) - n(\infty) = \frac{2}{\pi}  \int_{0}^{\infty} \frac{\lambda k(\lambda)}{1-\frac{\lambda^2}{\lambda^2_i}} \,d\lambda
\end{equation}

\noindent where we take $n(\infty)$ to be 0, the known real refractive index of each polymorph at a wavelength $\lambda_{\rm{i}}$ = 0.55 $\mu$m to be $n(\lambda_{\rm{i}})$ (see Table \ref{tab:material_props}), and $k(\lambda)$ to be the imaginary refractive index we computed from the laboratory data discussed above \citep{lippincott1958infrared, Philipp1985,Sitzars2000_sio2_ir_spectra,Koike2013}.  Given the form of the integral, we require the wavelength grid to be very finely spaced, at constant discretisation, and significantly wider than the range of integration. Therefore we interpolate our {\tridy} and {\cristo} $k$ values to a grid of $\lambda$ from 0.02 to 35 $\mu$m in steps of 3.5 $\times 10^{-3}$ $\mu$m, which we then bin down to the resolution of the quartz and amorphous silica refractive indices used in \citetalias{Grant2023clouds}. Our n values have corresponding uncertainty with k based on the Kramers-Kronig relation. Clearly, better laboratory data with fewer unknowns are required for silica polymorphs.

With the complex refractive indices in hand (subject to the significant uncertainties discussed above), we compute Mie coefficients for each polymorph at a given particle size using \texttt{PyMieScatt}'s \texttt{MieQCoreShell} routine \citep{piemiescatt} over the standard {\virga} particle size grid \citep{Batalha2020}. This grid ranges from 1$\times$10$^{-8}$ cm to 5.4 $\times$10$^{-2}$ cm in 60 steps. Our literature and calculated complex refractive indices, and an example of the extinction efficiency $Q_{\rm{ext}}$ calculated for a 32 nm radius particle, is presented in Figure \ref{fig:qext}. The refractive indices and Mie efficiencies clearly diverge between the different silica polymorphs, which we next input into the coupled {\picaso}-{\virga} framework described below.

\section{Atmospheric Models with {\picaso} and {\virga}}\label{sec:picaso}

To generate cloudy hot Jupiter atmospheric models, we use the {\picaso} 3.0 climate and radiative transfer code \citep{Batalha2019,Mukherjee2023} coupled to the cloud model {\virga} \citep{Batalha2020,Rooney2022}. {\virga} is the Python implementation of the \citet{ackerman2001} \texttt{eddysed} approach, which balances vertical mixing (parametrized by the eddy diffusivity, $K_{\rm{zz}}$) against particle rain-out (parametrized by a sedimentation efficiency factor, $f_{\rm{sed}}$). For the L dwarf atmospheric model, we use the \texttt{Sonora} \citep{Marley2021} grid to obtain a baseline temperature-pressure and chemical profile before adding post-processed {\virga} clouds and carrying out radiative transfer with {\picaso}.

\subsection{WASP-17 b Models}
Since \citetalias{Grant2023clouds} used {\picaso} and {\virga} models in their interpretation of WASP-17 b's atmosphere, we elect to use their best-fit chemistry, temperature-pressure profile, planetary, and cloud parameters. As such, for our WASP-17 b model, we use planetary parameters of 0.477 M$_{\rm{Jup}}$, 1.932~R$_{\rm{Jup}}$, a planetary equilibrium temperature of 1771 K, a stellar effective temperature of 6550 K, a stellar metallicity of -0.25, and a stellar log(g) of 4.149 \citep{Anderson2011,Southworth2012}. We also use \citetalias{Grant2023clouds}'s best-fit atmospheric profile of 81$\times$ solar metallicity, internal temperature 220 K, heat redistribution 0.67, and C/O ratio of 0.6 to include the atmospheric abundances of each molecule as can be accessed via Zenodo\footnote{https://doi.org/10.5281/zenodo.8360121}.

For {\picaso}'s radiative transfer, we account for opacities from \ce{CH4}, CO, \ce{CO2}, Cs, \ce{H2O}, \ce{H2S}, K, Li, \ce{N2O}, \ce{NH3}, Na, \ce{O2}, \ce{O3}, \ce{PH3}, Rb, TiO, and VO from 0.3--14 $\mu$m. Our opacities use the standard release {\picaso} v2 database\footnote{https://zenodo.org/records/3759675}, which is resampled to R=10,000 from an original R$\sim$10$^{6}$ line-by-line calculation \citep{Freedman2008}, appropriate for R=100 models. Our opacities are therefore slightly lower resolution (R=10,000 vs. R=60,000) and lack several species compared to that used in \citetalias{Grant2023clouds}, but we verify that we produce model spectra in reasonable agreement to those presented in \citetalias{Grant2023clouds}. These models are also of equal or higher resolution to the JWST/MIRI Low Resolution Spectrometer (LRS; \citealt{Kendrew2015}) data presented in \citetalias{Grant2023clouds}.

For the condensation of \ce{SiO2}, {\virga} uses the same expression as \citetalias{Grant2023clouds}, which is:

\begin{equation}
    10^4/T_{\rm{cond}} \approx 6.14 - 0.35(\rm{log}P_T) - 0.70[Fe/H]
\end{equation}

\noindent where $P_T$ is in bars, $T_{\rm{cond}}$ is the temperature in Kelvin, and [Fe/H] is the log of the atmospheric metallicity. Figure \ref{fig:phasediag} shows the temperature-pressure profile of WASP-17 b, along with the condensation curve of \ce{SiO2} following this expression for several atmospheric metallcities, demonstrating that higher metallicity shifts the curve to higher temperatures. The condensation curve of \ce{SiO2} droplets should be the same, regardless of what polymorphs ultimately form, though the nucleation energies required to condense directly from the gas phase to a solid particle may differ given the differences in mass density (Table \ref{tab:material_props}) and possible differences in surface energy and contact angle that arise from the different crystal arrangements \citep[see, e.g.,][for a discussion of conensate nucleation energy effects]{Gao2020}. {\virga} assumes that all possible condensing material condenses once the temperature-pressure profile crosses the saturation vapor pressure curve, and thus we ignore nucleation energy differences in our consideration of silica polymorph clouds. We encourage future microphysical modeling efforts to explore these effects.

From the material condensation curve and planetary temperature-pressure profile, {\virga} computes a log-normal particle size distribution of condensate scaled by the mass distribution of particles. In both \citetalias{Grant2023clouds} and this work, we use a non-typical {\virga} log-normal width of 1.2, which tightens the spread of possible particle sizes. A narrow range of particle sizes is required for a sharp cloud absorption feature to emerge from the spectrum.  This particle size distribution is then used to generate Mie coefficients for each atmospheric layer where condensate is present. From these parameters, {\virga} then outputs condensate optical depth, single scattering albedo, and asymmetry factors as a function of pressure and wavelength. These are input into {\picaso} to generate model transmission spectra.

Because {\virga} relies on mass balance to arrive at particle size distributions, the density of any condensate is a critical factor. The densities of each silica polymorph vary slightly as the specific crystal arrangement results in more or less dense atomic packing (see Table \ref{tab:material_props}). Built into {\virga} for \ce{SiO2} is 2.65 g cm$^{-3}$, which is that of $\alpha$-quartz. To explore the effect of polymorph density on the altitude and opacity of the silica cloud layer, we first generate model transmission spectra using the $\alpha$-quartz density for all polymorphs and only modify each model with appropriate optical properties. Next, we recompute each model allowing for both the correct optical properties \textit{and} density of each polymorph as shown in Table \ref{tab:material_props}. The results of both sets of model spectra are shown in Figure \ref{fig:w17spec}.

While we focus here on the silica cloud layer, we also found it necessary to include a lower \ce{Al2O3} cloud to achieve good fits to the data using the best-fit cloud mixing parameter values of $K_{\rm{zz}}$ (10$^{9.28}$) and $f_{\rm{sed}}$ (0.322) of \citet{Grant2023clouds}. \citetalias{Grant2023clouds} only briefly mentions the lower \ce{Al2O3} cloud layer, as it does not impact the longer infrared JWST/MIRI observations where the silicate feature dominates the spectrum. However, we find that inclusion of the \ce{Al2O3} cloud layer is critical to replicate the optical scattering slope observed by Hubble \citep{Alderson2022}, as seen in Figure \ref{fig:w17spec}. With more flexible atmospheric retrieval approaches using \texttt{petitRADTRANS} and \texttt{POSEIDON}, \citetalias{Grant2023clouds} did not require these \ce{Al2O3} clouds. However, since our analysis uses the forward {\virga} model alone, we include \ce{Al2O3} clouds in our baseline models for WASP-17~b.

To determine the best-fit polymorph cloud model compared to the observational data, we rebin our synthetic spectra to the resolution of the data presented in \citetalias{Grant2023clouds}, using both the Hubble data and Spitzer data \citep{Alderson2022} and offset included by \citetalias{Grant2023clouds}. We compute best fits by calculating the $\chi^2_{\nu}$ between the data and each forward model. We use the same offset between the relative transit depth and the data as in \citetalias{Grant2023clouds}. To assess the rank of our models, we compute the Bayesian Information Criterion (BIC) and the $\Delta$BIC following \citet{KassRaftery1995}:

\begin{equation}
    {\rm{BIC}} = \chi^2 + k ln(n) 
\end{equation}

\noindent where $k$ is the number of model parameters (in this case, either 1 for the optical properties or 2 for the optical properties and density), and $n$ is the number of data points, which is 95 for the combined HST/Spitzer/JWST dataset and 28 for JWST/MIRI LRS alone. The significance of $\Delta$BIC follows the intervals: 2 $<$ $\Delta$BIC as insignificant, 2 $<$ $\Delta$BIC $<$ 6 as positive, 6 $<$ $\Delta$BIC $<$ 10 as strong, and $\Delta$BIC $>$ 10 as very strong.

\subsection{L Dwarf Models}
To demonstrate the importance of considering silica polymorphs for brown dwarf atmospheres in addition to hot Jupiters, we also compute a series of forward models for each polymorph form in a putative L dwarf atmosphere. Following \citet{Burningham2021}, we use a \texttt{Sonora-Bobcat} model base appropriate for an L4.5 dwarf, with an effective temperature of 1700 K and a log(g) of 5.0, with a solar metallicity and C/O ratio. This temperature-pressure profile is plotted for reference in gold in Figure \ref{fig:phasediag}. The chemical abundances of this model can be found in the open source release on Zenodo\footnote{https://zenodo.org/records/5063476}. For these models, we use the same opacity database as in our WASP-17 b models, rebinned to R=3000, approximately that of JWST's MIRI Medium Resolution Spectrograph (MRS) across the wavelength range from 4.9 to 14 $\mu$m. 

We use a nominal $K_{\rm{zz}}$ of 10$^5$, an f$_{\rm{sed}}$ of 1, and a log-normal width of 1.2 to compute our {\virga} post-processed clouds. The width of this log-normal particle size distribution is significantly lower than a standard \texttt{Sonora}-{\virga} run. As noted by \citet{Burningham2021}, the standard \texttt{eddysed} (or in our case, {\virga}) scheme computes too wide a spread of particle sizes in distinct layers that blend out specific cloud features in favor of a broader, muddled silicate band that does not, at least in the case of the L4.5 dwarf 2MASSW J2224438-
015852 \citep{Burningham2021}, match observations. Therefore, we tune this parameter to a lower value to more clearly demonstrate differences that may arise from varied silica polymorphs. We do not attempt to fit or interpret the goodness-of-fit of these forward model runs, but merely perform them as a proof-of-concept for follow-up studies.

\section{Comparisons to WASP-17 b} \label{sec:w17b}

\begin{figure*}[t!]
\centering
{\includegraphics[clip, trim=0.cm 8.cm 0.cm 0cm,width=0.99\textwidth]{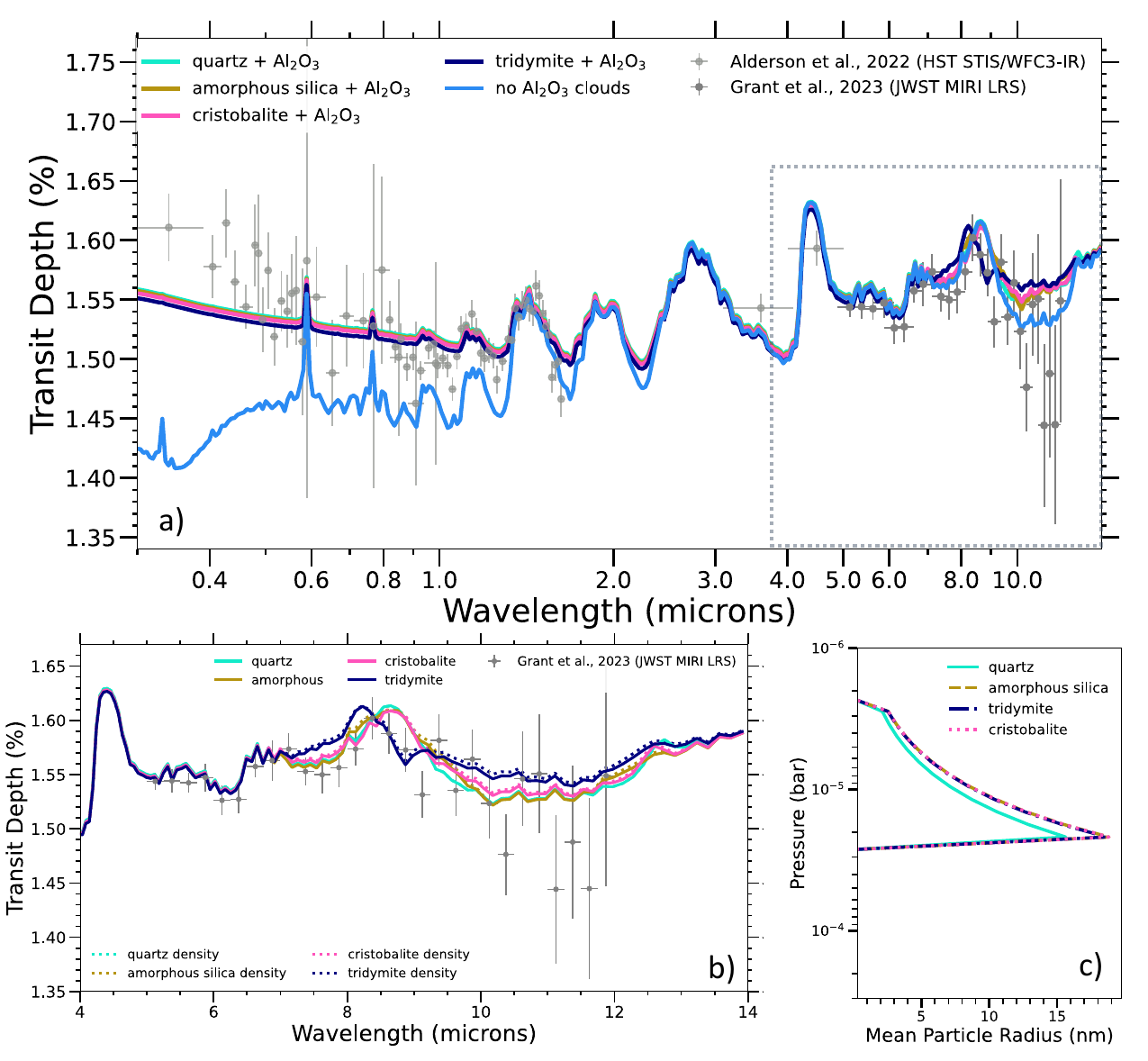}}
\caption{a) Atmospheric models of WASP-17 b using the Mie coefficients of  different silica polymorphs, using the best-fit sedimentation efficiency and eddy diffusivity of the \texttt{PICASO-Virga} nested sampling analysis of \citet{Grant2023clouds}. Shown in light blue is the spectrum without the contribution from \ce{Al2O3} clouds. b) The same as (a), but focused on the MIRI/LRS wavelength region as highlighted in (a) by the dotted grey box. Atmospheric models that account for the appropriate polymorph density in addition to optical properties are shown as dotted lines. c) Particle size distributions for each SiO$_2$ polymorph using appropriate polymorph densities.
\textbf{Accounting for silica polymorphs result in differentiable effects on the observable transmission spectrum of WASP-17 b.}}
\label{fig:w17spec}
\end{figure*}

Our results show that accounting for individual \ce{SiO2} polymorph optical properties does have demonstrable effects on the resulting planetary transmission spectrum. Figure \ref{fig:w17spec} shows the outputs of our model runs. We provide a complete breakdown of our statistical fits to each model in the Appendix in Table \ref{tab:stats} for the full data range and in Table \ref{tab:stats_miri_only} for the MIRI only data. Overall, we compute marginal differences in {\chisq} (i.e., $\chi^2/n$ where $n$ is the number of data points) between models with differing polymorph optical properties, similar to what \citetalias{Grant2023clouds} found between amorphous silica clouds and quartz clouds, where they reported a {\chisq} of 1.05 and 0.98, respectively. Our amorphous and quartz cloud fits have {\chisq} of 1.33 each, rather than finding a slight preference for quartz as \citetalias{Grant2023clouds} did. These differences in {\chisq} most likely stem from the lower resolution (R=10,000)  of our gaseous opacity database compared to \citetalias{Grant2023clouds} (R=60,000). However, as represented by the $\Delta$BIC, both our sets of models and \citetalias{Grant2023clouds}'s are within $\Delta$BIC of 0.3--0.6 of each other, which is not a significant interval. The exception is tridymite, which has a $\Delta$BIC  of 5.3 from the best-fitting model, a positive interval suggesting the other polymorphs are preferred. We discuss this poor tridymite fit in more detail below.

\subsection{Silica Polymorph Optical Properties}

Examining the JWST MIRI/LRS region in panel b of Figure \ref{fig:w17spec}, the differences between polymorph optical properties, even with the same set of cloud mixing values, is apparent by eye. The tridymite clouds shift the Si--O peak distinctly towards bluer wavelengths, with the cloud opacity better capturing the MIRI/LRS data peak at 8.38 $\mu$m and sharp dip in opacity around 9.12 $\mu$m. However, the tridymite has too much opacity to fit the data well from 7.1 to 8.1 $\mu$m, resulting in an overall {\chisq} of 1.38 for tridymite compared to 1.33 for quartz. We note that the error and extrapolation inherent to the overlapping, non-compatible datasets \citep{lippincott1958infrared,Sitzars2000_sio2_ir_spectra} used to construct the tridymite refractive indices make the uncertainty in the region from 6.7 -- 9.0 $\mu$m particularly high. The Appendix contains an extended run using an alternative version of tridymite optical properties (Figure \ref{fig:tridymite_optprops}). Indeed, we find that the region from 7.3 to 8.1 $\mu$m drives the relatively poorer fit for the tridymite clouds. Removing this region from the tridymite fit produces a {\chisq} of 1.317, which is the best-fitting model overall with the nominal cloud mixing {\fsed} and {\kzz} values. Adjusting for this poorly fit region also brings the $\Delta$BIC value for tridymite to a non-significant interval compared to the other polymorph fits, suggesting statistically equally good fits between them.

Cristobalite has a narrower blue and red edge for the Si--O peak, compared to either quartz or amorphous silica. Cristobalite clouds also have slightly increased opacity tapering off toward the red edge of the feature. The narrower blue edge of the quartz feature, along with with the slightly bluer peak overall compared to amorphous silica, is primarily what drove \citetalias{Grant2023clouds}'s preference for quartz over silica in the first place. With cristobalite, the subtle shift in opacity produces a {\chisq} of 1.32, which is better (though again, not to strong statistical significance) compared to either quartz or amorphous silica.

Either tridymite or cristobalite clouds would be self-consistent with the temperature of the atmosphere at this altitude, depending on the exact pressure level of cloud nucleation. Some combination of both phases is also possible. Laboratory data that is continuous over the MIRI wavelength region, at exoplanetary temperatures of $\sim$1300 K -- 2000 K, are needed to truly access the compatibility of the tridymite and cristobalite fits for WASP-17 b, as well as for future studies of exoplanetary atmospheric clouds.

\subsection{Silica Polymorph Optical Properties and Densities}

The addition of accurate densities only very subtly changes the particle distributions obtained from the nominal cloud mixing parameters, and thus the resulting planetary transmission spectra are similarly only subtly altered. The dotted lines of Figure \ref{fig:w17spec}b show the effect of proper density runs for each polymorph on top of the corresponding optical properties. The density of quartz is 2.65 g cm$^{-3}$  compared to $\sim$2.2 g cm$^{-3}$ for $\beta$-cristobalite, $\beta$-tridymite, and amorphous silica. Therefore the maximum mean particle size changes from $\sim$15 nanometers for quartz to $\sim$19 nanometers for the lower density polymorphs, as seen in Figure \ref{fig:w17spec}c. 

The statistical fits when adding density variance shift by {\chisq}s of only 0.005 to 0.007 for cristobalite and amorphous silica, and by a {\chisq} of 0.01 for tridymite. When comparing $\Delta$BICs against the non-density variation runs, this results in a positive but not strong preference for the uniform density due to the reduction of model parameters. Comparing the $\Delta$BICs of only the density runs, we see no significant difference in the fits, except for tridymite, which again is driven by the poorly fit extrapolated region from 7.3 to 8.1 $\mu$m. There may be outlier parameter space of highly extended cloud layers or low gravity planets where these density differences are more significant, but we leave this exploration to future work.

\subsection{Sedimentation Efficiency Variations}

Figure \ref{fig:w17spec}a showcases the complete spectrum as computed using the same planetary and cloud mixing parameters as in the original observational paper \citep{Grant2023clouds}. We show in light blue that, using quartz cloud opacity with the best-fit {\fsed} and {\kzz} of \citetalias{Grant2023clouds}, a deeper \ce{Al2O3} cloud is necessary to recover the optical-to-NIR scattering slope close to the Hubble data of \citet{Alderson2022}. Without this lower \ce{Al2O3} cloud, the fits for each polymorph fall to over {\chisq} $\sim$ 5.5 compared to {\chisq} $\sim$ 1.3. 

As part of our exploration of the model parameter space, we included several {\picaso}/{\virga} runs varying the sedimentation efficiency $f_{\rm{sed}}$ but maintaining the eddy diffusivity {\kzz}. We find an interesting solution wherein a lower level \ce{Al2O3} cloud layer is not needed to match the JWST, HST, and Spitzer observations of WASP-17 b presented by \citet{Grant2023clouds}. Their {\picaso}/{\virga} winning model required this lower alumina cloud with their best-fit $f_{\rm{sed}}$ of 0.3. However, we find that by tuning $f_{\rm{sed}}$ to higher values -- e.g., up to 3 -- we can generate a model that fits with slightly worse $\chi^2_{\nu}$ though statistically unfavored BIC values with silica clouds alone (see Table \ref{tab:stats} in the Appendix). This larger $f_{\rm{sed}}$ results in mean particle sizes for the cloud deck that are an order of magnitude larger -- up to 190 nanometers (Figure \ref{fig:fsed} in the Appendix, panel b). The cloud deck is slightly more compact as well, extending 2.4 $\mu$bar less than the fiducial \citetalias{Grant2023clouds} cloud. Such large $f_{\rm{sed}}$ is unexpected for hot Jupiters, which are typically inferred to have $f_{\rm{sed}}$ less than 1 \citep[e.g.,][]{ackerman2001,morley2015}.

While the \texttt{POSEIDON} and \texttt{petitRADTRANS} retrievals performed in  \citet{Grant2023clouds} were also able to find solutions without an alumina cloud layer at depth, they still found mean particle sizes on the order of 10 -- 20 nanometers. We tested all polymorphs in this model parameter sweep, and found consistent results that this high {\fsed} case can match the data without the need for the \ce{Al2O3} cloud at depth for each silica phase. We show only tridymite in Figure \ref{fig:fsed} for simplicity. Moreover, if we tune our $f_{\rm{sed}}$ up to 3 but keep the \ce{Al2O3} cloud deck, we find improved fits even beyond that of the nominal $f_{\rm{sed}}$ = 0.3 run, with similarly larger particles. These fits are positively to strongly preferred over the nominal runs according to their BIC values. These results highlight that model searches beyond expected values for tuning parameters can offer unique insights into the structure of exoplanetary atmospheres. 

\begin{figure*}[t!]
\centering
{\includegraphics[width=0.99\textwidth]{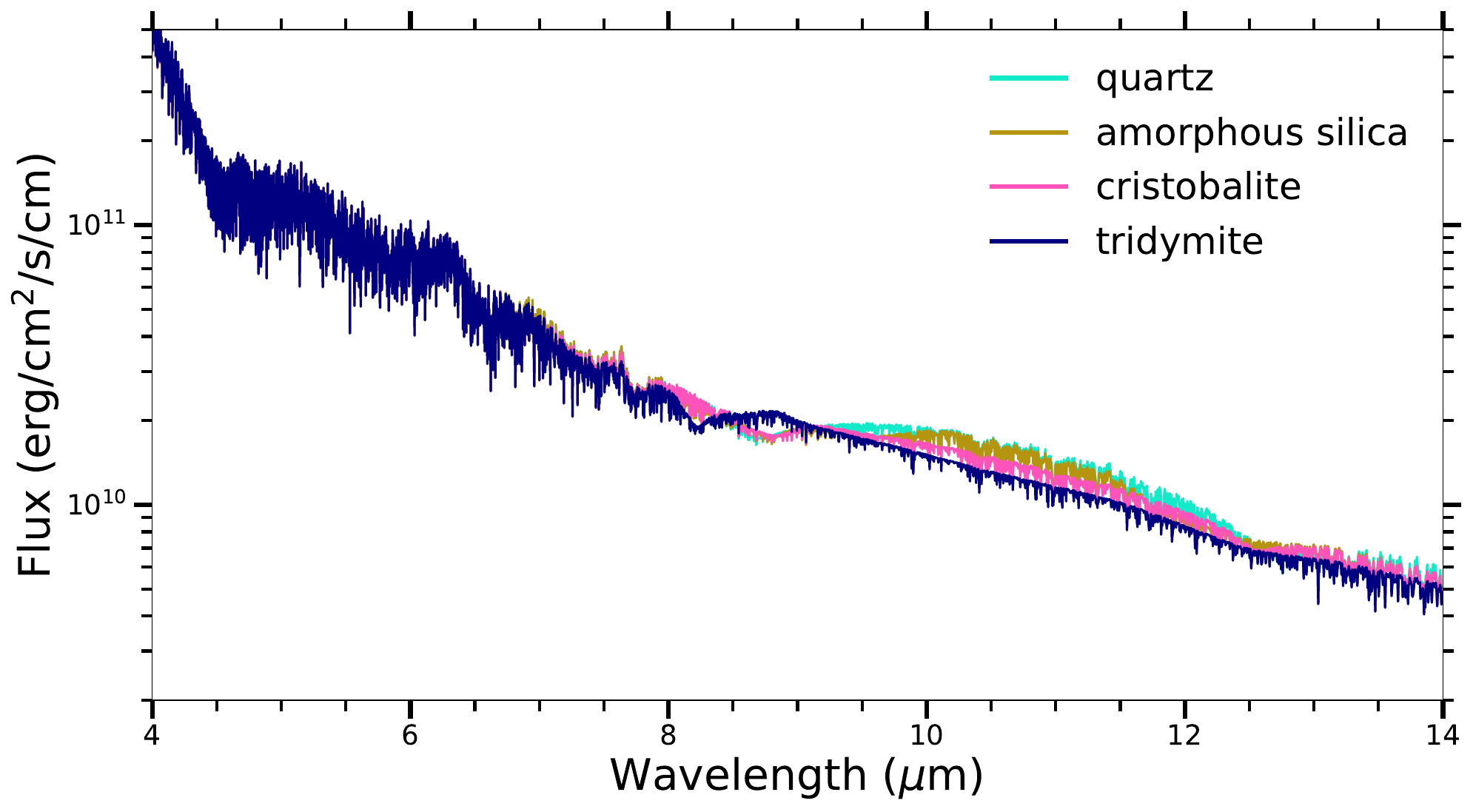}}
\caption{Emission spectra for \texttt{Sonora-Bobcat} brown dwarf atmospheric models with effective temperature of 1700 K and surface gravity log(g) = 5.0, post-processed with clouds generated with different Mie coefficients for silica polymorph cloud particles.  \textbf{In addition to affecting planetary transmission spectra, different polymorphs of crystalline silica clouds could be also differentiable with JWST/MIRI in emission for a wide range of warm substellar objects.}}
\label{fig:bd_emission}
\end{figure*}

\section{Predictions for Silica Polymorphs in L Dwarf Atmospheres} \label{sec:BD}

Our brown dwarf {\sonorabob} models post-processed with the optical properties of the four silica polymorph phases are shown in Figure \ref{fig:bd_emission}. Again, as with their exoplanet counterparts, we note visible differences in the spectra. The sharper, bluer tridymite cloud Si--O feature is strongest just short of 8.4 $\mu$m, compared to the quartz, cristobalite, and amorphous silica cloud features which are strongest just short of 9 $\mu$m. At redder wavelengths from 9 to 12 $\mu$m, the emergent flux from the quartz and amorphous silica models is stronger compared to both tridymite and cristobalite due to the higher extinction of the latter (as was demonstrated in Figure \ref{fig:qext}.) 

We do not attempt to fit any observational data to these model runs, as this exercise is meant primarily as motivation for considering mineral cloud polymorphs in follow-up studies both theoretical and observational for brown dwarfs. As noted in \citet{Burningham2021} and others, some combination of magnesium silicates, silica, and other refractory clouds could all contribute to the generally wide ``silicate index'' \citep{Suarez2022} observed in L dwarfs. As in Figure \ref{fig:phasediag}, the temperature-pressure profile of a typical L dwarf crosses through the stability regions for all the silica polymorphs at 0.1 bar to millibar pressures, so the exact point of nucleation will matter for the formation of any particular phase cloud particle.  Consideration of multiple silica polymorphs can help contribute to this broadening in the ``silicate index'' region without invoking additional cloud species, which should be carefully explored in future L dwarf studies. 

\section{Discussion and Future Work} \label{sec:disc_concl}

\subsection{Differentiating Silica Polymorphs}

As we have shown in Figures \ref{fig:w17spec} and \ref{fig:bd_emission}, the silica polymorphs can be distinguished by eye in atmospheric models in both emission and transmission for brown dwarfs and hot Jupiters. However, given the current data quality for WASP-17~b, only the tridymite clouds can be strongly statistically differentiated with current signal-to-noise (SNR). On the other hand, due to both higher SNR and resolution from brown dwarfs and directly imaged planetary emission data, all silica polymorphs should be distinguishable should individual features be present (see the data quality, e.g., of VHS-1256~b; \citealt{Miles2023}). To statistically differentiate between cristobalite and quartz clouds for WASP-17~b would require an SNR at the current R$\sim$100 of 300--500 compared to the current SNR$\sim$75 from 7 to 11 $\mu$m. Such SNR could potentially be achievable for some hot Jupiter targets, including of WASP-17~b, by stacking multiple transit observations. Alternatively, higher resolution observations could be performed in transit with MIRI's Medium Resolution Spectrometer, with R$\sim$3000 over wavelengths from 5 to 12 $\mu$m. However, the signal-to-noise and general feasbility for such observations for this mode is as of yet uncharacterized in transit \citep{Deming2024}.

In addition to SNR and resolution considerations, disentangling polymorph cloud signatures from each other is dependent on whether observations are performed in emission or transmission, as these are probing different atmospheric regions. In transmission only the limb of the planet is in view, while in emission the dayside of the planet is accessible. For a brown dwarf, emission gives information on the entire disk of the object and can vary with rotation \citep[][]{Biller2024}. Transmission measurements also probe at slant geometry which can enhance the optical depth of clouds at high altitudes and low (millibar) pressures \citep[e.g.,][]{Fortney2005}, while brown dwarf emission measurements are probing down to (bar) pressures nearing the photosphere \citep[e.g.,][]{Brock2021}. The combination of the differing viewing geometry and contribution functions means that different cloud layers -- with varying particle size distributions, cloud coverage, and competition from varying gas opacities -- are being probed with the differing observational techniques. Whether silica cloud polymorphs will be resolvable is thus a balance between all these effects. The smallest particle sizes -- with the strongest distinctions in polymorph features -- may be accessible more easily with transmission spectroscopy that probes higher in the atmosphere, while higher SNR and resolution for directly imaged planets and brown dwarfs may allow the identification of polymorphs even when deeper, larger particle cloud decks are being probed.

\subsection{Laboratory Data at Relevant Conditions}

Given the large uncertainties due to our necessary extrapolations for the refractive indices, all our ``best-fit'' models should be taken with an enormous grain of salt\footnote{or sand, if you will}. Our purpose in this work is to demonstrate that consideration of the observational effects of polymorphs is warranted. Crucially, extensive laboratory data is required to hone in on the suggested trends and inferred particle properties we introduce by including these polymorphs in our modeling analysis.

As highlighted by \citet{Potapov&Bouwman2022}, temperature and pressure are critical controls on the ultimate spectra of astrophysical materials, and the values we must use here fall short. Moreover, since no reliable data exist for these polymorphs at ultraviolet to optical to near-infrared wavelengths, we could be missing critical information that could further differentiate these structures and provide further constraints on atmospheric conditions.  Notably, previous high temperature laboratory evidence for the formation of tridymite and cristobalite in astrophysical environments relies on X-ray diffraction crystallography \citep{Fabian2000}, which should likely be performed in conjunction with spectral measurements to ascertain both the structure and spectral impact of different polymorphs. 

As noted in Section \ref{sec:polymorphs}, a complete lack of temperature-pressure stability constraints exists for high temperature polymorphs below 1 bar in pressure, which is crucial to interpreting the ultimate fate of these particles as clouds in substellar atmospheres. Moreover, the timescales for these polymorphs to undergo phase transitions between the various crystalline arrangements has also, to our knowledge, never been measured at these elevated temperatures and pressures in the melt $\rightarrow$ solid direction of decreasing temperature. Instead, data exist regarding annealing from low to high temperature, or as shock-quenched glasses. As some crystallization experiments suggest that the structural evolution of silica polymorphs on short timescales can be strongly influenced by their initial structures \citep{HillRoy1958}, both directions of thermal evolution, at low pressures, must be carefully measured to fully investigate the impact of a particle's trajectory throughout a substellar atmosphere. 

\subsection{Beyond Mie Theory}

In our computation of cloud opacity, we have used Mie Theory, which inherently assumes particles are spheres, as does most of the exoplanet cloud literature for computational speed. Given the crystalline nature of silica polymorphs, this spherical assumption is necessarily incorrect. Exoplanet atmospheric studies are increasingly including consideration of non-spherical aggregate cloud particles, which will change not only the opacities of such material \citep[e.g.,][]{Min2006,Min2015,dominik2021,ohno2020,lodge2024,vahidinia2024}, but also their sedimentation and lofting throughout the atmosphere \citep{Adams2019,ohno2020,Samra2020,Samra2022,vahidinia2024}. Here, the differing densities of polymorphs, in addition to their unique infrared absorption, will need to be accounted for. 

A further consideration is the growth mechanism of crystalline aggregates. The mathematical representation of aggregate particles must be informed by the primary method of growth \citep[e.g.,][]{ohno2020}. For example, cristobalite has been experimentally observed to grow spherulitically, while quartz and tridymite grow as oriented crystalline films \citep{Guinel2006}. Partial crystallization will add an additional layer of complexity to accurately describing cloud particle shapes if modeling more than simple spheres. We encourage future studies along this line of inquiry, which may add additional observables to unravel the complexity of polymorphic clouds.

\subsection{Disequilibrium and Intermediary Cloud Compositions}

The presence of alkali metals can also play a critical role in the formation of different SiO$_2$ polymorphs. High temperature annealing experiments with SiO$_2$ grains demonstrate that formation of tridymite requires minor concentrations ($<$ 1 wt\%) of ``mineralizing agents" like Na or K \citep{MosesmanPitzer1941, DeSousa2014, Dapiaggi2015}. Counter-intuitively, the presence of alkali metals causes SiO$_2$ to partially crystallize first as cristobalite before reverting to the more a stable tridymite structure, even when the temperature is below the nominal stability range of pure cristobalite. Formation of cristobalite in these conditions most likely occurs while SiO$_2$ passes through a transient amorphous stage during transformation between polymorphs. The crystallization rate of tridymite through this transitory phase occurs on a timescale of several hours at 1273 K in the presence of Na, with up to 60 wt\% of SiO$_2$ transforming to tridymite within 6 hours during annealing experiments with SiO$_2$ grains $<$30 $\mu$m \citep{Dapiaggi2015}. The same transformational sequence is observed with the presence of minor K, but at a much more sluggish rate. 

Thus, the presence of alkali metals permits the coexistence of cristobalite and tridymite in SiO$_2$ grains if the cooling rate is faster than the rate of polymorphic transformation. This may be the case for exoplanetary cloud particles lofted from higher temperature regions at greater depths in the atmosphere. Extending this line of reasoning, the co-presence of both silica in addition to magnesium silicate cloud particles, as well as intermediary phases containing Fe, as inferred in previous studies \citep[e.g.,][]{Burningham2021},  may also serve to broaden and alter the shape of the silicate cloud feature in substellar atmospheres. 

\subsection{Liquid, Crystal, or Amorphous Cloud Particles}

Whether or not silica clouds initially form as liquid droplets, crystalline ``snowflakes'', or amorphous glassy particles has major implications for the clouds ultimately observed by telescopes. For higher temperature objects, such as the L dwarf and WASP-17 b profiles shown in Figure \ref{fig:phasediag}, the \ce{SiO2} condensation curve and the T-P profiles cross at 0.1 bar to millibar pressures at temperatures very near the melt-to-cristobalite transition. Slight differences in atmospheric metallicity or nucleation energy make it likely reasonable to assume these clouds form as crystalline particles, whereupon long (i.e., geologic) timescales are required for them to relax down to a lower polymorph phase even if they are subsequently moved to cooler regions of the atmosphere.

On the other hand, significantly cooler objects that have observed silica or silicate cloud features challenge this assumption. An example is that of the highly inflated WASP-107 b, which has an equilibrium temperature of only 750 K, 1000 K less than our case study objects of WASP-17 b and the L4.5 dwarf. Recent JWST observations have shown WASP-107 b requires a very high internal temperature of 300--500 K which is potentially driven by tidal heating \citep{Welbanks2024}. This internal heat flux is several hundred K higher than would be expected given the size and age of the planet \citep{Dyrek2024,Sing2024,Welbanks2024}. 

JWST MIRI/LRS observations (5 -- 12 $\mu$m), combined with past HST Wide Field Camera 3 (WFC3) data (0.8 -- 1.6 $\mu$m; \citealt{Kreidberg2018,Spake2018}) strongly (to 7$\sigma$) require the presence of silicate clouds to explain a spectral feature around 10 $\mu$m, muted water features, and the near-infrared scattering slope. The authors included a mixture of amorphous SiO, \ce{SiO2}, and \ce{MgSiO3} particles at millibar pressure levels \citep{Dyrek2024}, though they did not test crystalline forms of these silicates. Another study combined NIRCam observations (2.4 -- 5 $\mu$m) with the previous MIRI/LRS and HST/WFC3 data (covering 0.8 to 12 $\mu$m, in total) and also found a millibar level silicate cloud base would be required to match the data \citep{Welbanks2024}, while a NIRSpec G395H spectrum (2.9 -- 5 $\mu$m) with depleted methane requires a very warm interior \citep{Sing2024}.

The presence of such a high silicate cloud layer, along with the absence of \ce{CH4}, caused all three studies to infer very high vertical mixing rates, with eddy diffusivities of $K_{\rm{zz}}$ = 10$^8$--10$^{12}$ cm$^{2}$ s$^{-1}$ \citep{Dyrek2024,Sing2024,Welbanks2024}. In this scenario, silica cloud particles would form at depth (10s to 100s of bar) and be lofted to observable millibar levels. Examining the teal curve for WASP-107 b in Figure \ref{fig:phasediag} (taken from \citealt{Dyrek2024}), we see that at these depths and temperatures, silica clouds would form as liquid droplets. It is therefore instructive to estimate how quickly such a liquid cloud droplet would rise through the atmosphere to the millibar pressures of the inferred cloud deck, including how long it spends in each polymorph stability region, to ascertain the likeliest silica polymorph for this object. 

We can approximate a vertical mixing timescale by relating the vertical eddy diffusion $K_{\rm{zz}}$ to the atmospheric scale height $H$ \citep[e.g.,][]{Komacek2019,powell2024twodimensional}:
\begin{equation}
    \tau_{\rm{dyn}} \approx \frac{H^2}{K_{\rm{zz}}}
\end{equation}

\noindent where the scale height is defined as:

\begin{equation}
    H = \frac{k_{\rm{B}}T(z)}{\mu g}.
\end{equation}.

\noindent Then, we can equate the pressure $P$ and the altitude $z$ using the standard relation with scale height:

\begin{equation}
    P(z) = P_{\rm{0}} e^{-z/H(z)}
\end{equation},

\noindent to find the number of scale heights traversed by the particle, where $P_{\rm{0}}$ is the pressure at which the particle initially condenses and $P(z)$ is the point at which the cloud particle is observed.

Using WASP-107 b as our example, we can see from Figure \ref{fig:phasediag} that the planet's pressure-temperature profile crosses the \ce{SiO2} condensation curve at approximately 30 bar. Along its ascent, the particle transitions through the stability regions of cristobalite, tridymite, and $\beta$-quartz before reaching the $\alpha$-quartz stability region at 1 millibar. Under the 10$\times$ metallicity inference of \citet{Dyrek2024}, the scale height of WASP-107 b is approximately 800 km, assuming a mean molecular weight $\mu$ of 2.8, an equilibrium temperature 750 K, and $g$ of 270 cm s$^{-2}$.  We'll assume here that the particle is coupled to the gas mixing timescale, for simplicity. Using the maximum inferred $\log K_{\rm{zz}} (\mathrm{cm^2\, s^{-1}})$ for this object, 11.7 \citep{Sing2024}, we can then estimate the dynamical timescale for a particle to be $\sim$ 1.3$\times$10$^4$ seconds, or 3 and a half hours for one scale height. The altitude change from 30 bars to a millibar is approximately 10 scale heights, so the total time it takes the particle -- moving at 220 km/hr -- to reach the observed cloud layer is on the order of 37 hours. 

This estimate is of course naively neglecting the effects of drag, non-convective atmospheric layers, or advection, which would all serve to alter this timescale \citep{Komacek2019}. Nevertheless, it provides a rough idea of the time for a silica particle to rise to observable levels. Thirty to forty hours is plenty of time for crystallization to occur, with the particle spending over 2 hours in the cristobalite region and 7 hours in the tridymite region, with some combination of tridymite and cristobalite particles thus likely for the observed cloud layer. Localized, much faster updrafts could perhaps cause the clouds to rise quickly enough to quench the liquid droplets into amorphous glassy silica. Some combination of polymorphs is clearly possible in a vigorously mixed atmosphere, which would all broaden the observed Si--O peak, though perhaps biased toward the polymorph that dominates the cloud mass. Accounting for the continued presence of cloud particles at these high altitudes, rather than having them rain out to below observable levels, is another question in the case of WASP-107 b \citep{Welbanks2024}.


\subsection{Silica Polymorph Clouds as  Meteorological Sensors}

In the previous subsection, we speculated on the vertical mixing of cloud particles and the implications for the fate of particular polymorph phases. However, substellar objects are 3-dimensional, which will impart more complicated dynamical mixing and advection in horizontal, longitudinal, and latitudinal directions.

A recent microphysical study showed that cloud formation efficiency and persistence is enhanced in a 2-dimensional framework, with certain cloud species able to be transported and survive on the daysides of
hot Jupiters in cases where 1-dimensional models would not predict the existence of clouds \citep{powell2024twodimensional}. Silicate clouds are thought to form readily on the nightsides of a wide range of hot Jupiters \citep{GaoPowell2021}, so accounting for the polymorph of silica clouds where they formed and their evolution to where they are observed has major diagnostic potential as a tracer of atmospheric thermal gradients, wind speeds, and rain-out. 

For brown dwarfs, observational studies combining decades of archival Spitzer data suggests that L dwarfs of L4-L6 spectral type are most silica-rich \citep{Suarez2022}, and that low-gravity, young atmospheres have broader, redder, silicate absorption. The authors take this as an indication of grain size and composition differences between condensates \citep{Suarez2023heaviergrains}. Consideration of silica polymorph cloud particles, with their distribution of Si--O peaks and differing densities, could further illuminate the dynamics of these objects. Finally, \citet{Suarez2023Latitudes} also suggest that equatorial regions are preferentially cloudier, which careful accounting of polymorph features and their stability regions could also help constrain. 

Because polymorphs record the thermal history of \ce{SiO2} grains, we propose that observing particular -- or multiple -- polymorph phases in silica cloud layers will act as atmospheric tracers relevant for a wide range of substellar atmospheres. General circulation models (GCMs) that include cloud tracers \citep[e.g.,][]{Roman2021,Lee2022,Steinrueck2023} coupled to phase-resolved observations \citep[e.g.,][]{LewisHammond2022,Hammond2024} accounting for cloud polymorph phase could offer major insight into the physical conditions of substellar objects.

\subsection{Applications for Lava Worlds and Silicate Vapor Atmospheres}

Here, we have focused on hot Jupiters and brown dwarfs, but the implications of silica polymorphs extend beyond gas giants. There is considerable recent interest in the idea of ``lava worlds,'' ultra-hot terrestrial planets which could have transient or tenuous silicate vapor atmospheres \citep[e.g.,][]{Zieba2022,Zilinskas2022,Piette2023,Falco2024,Hu2024}. If nightsides of these objects are cool enough, silicate clouds, potentially including silica if the chemistry of the outgassed atmosphere is favorable, could also form on these objects. If the atmosphere is escaping, the silicate atmosphere will also condense into dusty outflows \citep{Booth2023,CamposEstrada2024}. Moreover, if the atmosphere is tenuous enough, the nightsides of these planets could even experience atmospheric collapse, where a thin veneer of silicate ``ice'' lies on the surface, which could be detectable by albedo \citep{Mansfield2019} or phase curve measurements \citep{Kreidberg2019LHS}. Observations and models accounting for the silica polymorph likely to be stable in each regime could provide a tracer of the thermal history of the planetary surface and atmosphere, in addition to offering insight on the conditions of material outgassed from the interior. 

\subsection{Mineral Cloud Polymorphs Beyond Silica}
Since many substellar atmospheric clouds are made of mineral species, the stable polymorph of each mineral at the relevant temperature-pressure condition must be considered. An extensive discussion of all potential polymorph clouds is beyond the scope of this work, but we briefly mention a few potentially important cloud species here. The species discussed below are expected to be the most dominant cloud masses for a variety of substellar temperatures based on their nucleation efficiencies \citep{Gao2020}.


The magnesium silicates, \ce{MgSiO3} and \ce{Mg2SiO4}, are known as their enstatite and forsterite crystalline polymorphs at Earth surface conditions, in addition to having been identified in a multitude of different physical environments, including circumstellar shells surrounding evolved and young stars, comets, protoplanetary disks, and meteorites \citep[e.g.,][]{Hanner1994,Jang2024ProtoPlaDisk}. For forsterite, known polymorphs are limited to high pressure (GPa, i.e., 10$^4$ bar) conditions \citep{Presnall1995,Miyahara2021} not relevant to cloud formation. However for enstatite, the phases clinoenstatite,  orthoenstatite, and protoenstatite are all stable at various high temperature ($>$800 -- 1200 K)/low pressure conditions \citep{ManualMineral,Presnall1995}. Their exact stability regions are controversial to the point of the Earth surface ambient phase not being fully settled \citep{enstatite2000}, as some of these polymorphs are very challenging to synthesize and stabilize in the laboratory \citep{shugo2022}. Protoenstatite has successfully been made experimentally and seems to be stable once formed, but infrared spectroscopy of all the enstatite crystalline phases remains incomplete \citep{Roskosz2011,Matsuno2012}. These polymorphs do exhibit significant peak shifts in Raman spectroscopy \citep{Roskosz2011,Kanzaki2017}, so could very well have similar peak shifts in infrared spectra if measured.


ZnS requires condensation nuclei to form efficiently \citep{gao2018}. Assuming it does form, ZnS has two polymorphs, sphalerite (cubic) and wurtzite (hexagonal). These two crystal forms have slightly differing band gap energies and thus differing UV-Vis aborption peaks \citep{Kole2012}. Wurtzite is the stable form above 1300 K, which is very near the condensation temperature at 10s to 100s of mbar atmospheric pressures at elevated ($\gtrsim$ 50$\times$ solar) atmospheric metallicity \citep{Morley2012}. Sphalerite may be the more common observed form given the temperatures where ZnS is expected to be one of the uppermost cloud layers, however \citep{morley2015,gao2018}. Substellar literature uses only the sphalerite form for optical properties of ZnS clouds \citep{Morley2012,WakefordSing2015,KitzmannHeng,gao2018,Batalha2020,Molliere2019,Lee2022}. The infrared absorption of ZnS is significantly weaker than other cloud species, however, making it likely difficult to observe with JWST/MIRI \citep{WakefordSing2015}. Nevertheless, given the UV-Vis differences \citep{McCloy2009EffectsOT,Kole2012}, further study of ZnS polymorphs could be of interest for Hubble WFC3/UVIS and STIS studies, or eventual Habitable Worlds Observatory studies.

While \ce{TiO2} particles may be largely hidden under an \ce{Al2O3} cloud layer \citep{Gao2020}, the most stable \ce{TiO2} polymorph across all temperatures is rutile \citep{rutile}. \citet{KitzmannHeng}'s compilation of optical properties, which is also sourced by \texttt{petitRADTRANS} \citep{Molliere2019}, instead reports values for anatase from \citet{zeidler2011}. Zeidler and coauthors do also report values for rutile, which is sometimes used in other substellar atmospheric codes, such as \texttt{gCMCRT} \citep{gCMCRTsoftware,Lee2022}. Anatase, while potentially more relevant for circumstellar dust regimes, readily transforms to rutile above 1200 K \citep{zeidler2011}, where \ce{TiO2} clouds are expected in substellar atmospheres \citep[e.g.,][]{Helling2006,Lee2016,Gao2020}. 

For alumina itself, a number of metastable transition crystalline phases exist that exhibit variance in their infrared spectral features \citep{Gangwar2015}. However, corundum ($\alpha$-\ce{Al2O3}) is the only stable polymorph, regardless of temperature \citep{levin1998}, and should therefore be the dominant form of \ce{Al2O3} cloud particles in warm substellar atmospheres. Substellar atmospheric works frequently use amorphous alumina, \citep{KitzmannHeng, Lee2022}, though crystalline forms are also used \citep{WakefordSing2015,Molliere2019}. We note that \citet{zeidler} reports temperature-dependent corundum refractive indices which we encourage for future studies regarding alumina clouds. 

\subsection{Polymorph Cloud Features in JWST MIRI/MRS}

In addition to the primary Si--O stretching band around 8 to 10 $\mu$m, silica polymorphs have additional absorption bands near 18 to 24 $\mu$m, as visible in Figure \ref{fig:qext}. This longer wavelength feature arises from Si--O--Si bending modes. Like the $\sim$10 $\mu$m stretching feature, the exact wavelength of the bending mode absorption peak differs between each polymorph due to the differing bond energies unique to each crystal structure. These features may thus offer an additional diagnostic by which to determine the identity of cloud particles. 

However, JWST's MIRI Medium Resolution Spectrometer (MRS) has not yet been proven out or approved for widespread transiting exoplanet studies. Moreover, MIRI/MRS Channel 4, covering 17.7 to 28 $\mu$m, has markedly lower resolution and throughput compared to the shorter wavelength channels \citep{Wells_2015,ArgyriouMIRI,LabianoMIRI}. Observations of brown dwarfs and directly imaged planets have borne out this precipitous drop in information content, with the drop in precision, resolution, and flux at these extended wavelengths resulting in studies treating MIRI/MRS Channel 4 spectra as a photometric point \citep[e.g.,][]{Miles2023}. Therefore, it is unlikely that JWST observations will be able to make use of the 18 to 24 $\mu$m bending mode of silica to distinguish between the various possible polymorphs that could make up warm substellar clouds. Future facilities should consider this possibility in their instrument design to further constrain the physics of cloud formation in exotic atmospheres.

\section{Summary and Conclusions} \label{sec:summary}

In this work, we have reintroduced the idea of polymorphs -- that is, specific crystalline arrangements of minerals based on thermodynamic stability -- into the exoplanetary and substellar literature. Polymorphs have long been considered by both Earth and planetary geologists, as well as by protoplanetary astrophysics \citep[e.g.,][]{Fabian2000,Koike2013}.

Here, we focused on silica polymorphs in particular. We gathered what sparse laboratory data exist to compile and calculate optical properties for four silica polymorphs -- quartz, amorphous silica, tridymite, and cristobalite -- that should be stable at the elevated temperatures and low pressures of substellar upper atmospheres. We performed case studies for an exoplanet, WASP-17 b, and an L dwarf to demonstrate the observable effects of accounting for silica polymorph optical properties. We found that the cloud opacities do in fact diverge for both transmission and emission spectra when silica polymorphs are considered individually. We note that tridymite's optical properties are particularly uncertain, which drives both the goodness-of-fit and differentiability of this polymorph in particular.

We then proposed several lines of inquiry for follow-up studies of both silica and other mineral cloud polymorphs. These include more sophisticated modeling like the inclusion of non-Mie theory for cloud particles, investigation of the diversity of microphysical processes between mineral phases, exploration of cloud dynamics through both 2-dimensional models and GCMs, more complex cloud compositions, and applications for high temperature worlds across the mass range.

Our major conclusion is that mineral cloud polymorphs will act as ``witness plates,'' or diagnostic tracers of thermal conditions throughout the atmosphere. Given the complexity of atmospheric dynamics, day-night temperature contrasts, cloud nucleation, and thermal structure, we expect that combinations of polymorphs are quite likely.

In summary, we have shown that the spectral effects of silica polymorphs are readily distinguishable within the resolution and wavelengths of JWST/MIRI. However, we urgently require laboratory measurements of these materials at sufficient resolution, wavelength coverage, and temperature/pressure conditions to be truly relevant and applicable for JWST studies of substellar atmospheres.  
Once equipped with adequate laboratory datasets, we recommend that modelers of these atmospheres no longer neglect mineral polymorphs.

\nolinenumbers{
\begin{acknowledgments}
The authors thank N.E.~Batalha for sharing the best-fit model for WASP-17 b, C.~Koike for sharing the cristobalite absorption data, and H.R.~Wakeford and M.J.~Radke for helpful discussions. S.E.~Moran thanks B.W.~Patterson for painstakingly manually extracting the tridymite transmission spectrum from the 1958 figure, nonsensical axis scale and all. Portions of this work were performed in support of observations made with NASA/ESA/CSA's JWST, which is operated by the Association of Universities for Research in Astronomy, Inc., under NASA contract NAS 5-03127 for JWST. This work was performed in part in support of observations associated with program \#2288.

\end{acknowledgments}
}
\software{
Astropy \citep{astropy,astropy2},  IPython \citep{ipython}, Matplotlib \citep{matplotlib}, \texttt{matplotlib-label-lines} (Cadiou et al., 2022; https://zenodo.org/record/7428071), NumPy \citep{numpy, numpynew},  \texttt{PICASO} \citep{Batalha2019,Mukherjee2023},
\texttt{pyElli} (Müller and Dobener; https://github.com/PyEllips/pyElli), pysynphot \citep{STScIDevelopmentTeam2013},
\texttt{PyMieScatt} \citep{piemiescatt},
\texttt{Sonora} \citep{Marley2021},
{\virga} \citep{Batalha2020,Rooney2022}}

\bibliography{polymorphs_revision2}{}
\bibliographystyle{aasjournal}



\appendix

\section{Statistical Fits of Polymorph Models to WASP-17 b data}
\begin{table}[h]
\centering
\begin{tabular}{@{}lllllll@{}}
\toprule
Model                                         & $\chi^2$  & $\chi^2/n$ & BIC & $\Delta$BIC  \\ \midrule
\textit{Mie coefficients, BIC k = 1, n = 95, dof = 94}                 &      &     &                   \\
quartz                                        & 126.3  &     1.330 & 130.9    &  0.6                 \\
amorphous silica                              &  126.0 &   1.327 & 130.6    &   0.3               \\
tridymite                                     & 131.0 &    1.379** &  135.6   &  5.3             \\
cristobalite                                  &  125.7 &    1.323 & 130.3    &   --                \\ \midrule
\textit{Mie coefficients and density, BIC k = 2, n = 95, dof = 93}    &      &     &                   \\
quartz                                        & 126.3 & 1.330    &  135.4   &   5.1                \\
amorphous silica                              & 125.4 & 1.320    &   134.5  &   4.2             \\
tridymite                                     &  131.9 & 1.389    &  141.0   &  10.1             \\
cristobalite                                  &  125.2 & 1.318   &  134.3   &  4.0              \\ \midrule \midrule \midrule
 \multicolumn{5}{c}{Supplemental Parameter Space Fits} \\
\textit{f$_{\rm{sed}}$ variations, BIC k = 1, n = 95, dof = 94}            &      &     &                   \\
quartz only, $f_{\rm{sed}}$ = 3               & 177.8 & 1.872     & 182.4    &  59.8                 \\ 
quartz only, $f_{\rm{sed}}$ = 0.3             & 525.0 & 5.526     &   529.6  &  407.0                 \\ 
quartz, $f_{\rm{sed}}$ = 3             & 118.1 & 1.243     &   122.6  &  --                 \\ 
\\
amorphous only, $f_{\rm{sed}}$ = 3               & 178.3 & 1.877     & 182.9    & 60.3                 \\ 
amorphous only, $f_{\rm{sed}}$ = 0.3             & 528.9 & 5.567     &   533.5  &  410.9                \\ 
amorphous, $f_{\rm{sed}}$ = 3             & 118.7 & 1.249     &   123.3  &  0.7                 \\ 
\\
tridymite only, $f_{\rm{sed}}$ = 3            & 198.0 & 2.084     & 202.6    &  80.0                \\ 
tridymite only, $f_{\rm{sed}}$ = 0.3          & 546.1 & 5.748     & 550.7  &  428.1              \\ 
tridymite, $f_{\rm{sed}}$ = 3          & 126.3 & 1.329     &  130.9  &  8.3     &         \\ 
\\
cristobalite only, $f_{\rm{sed}}$ = 3               & 187.7 & 1.976     & 192.3    &  69.7                \\ 
cristobalite only, $f_{\rm{sed}}$ = 0.3             & 529.8 & 5.577     &  534.4  &  411.8                \\ 
cristobalite, $f_{\rm{sed}}$ = 3             & 120.1 & 1.264     &  124.7  &  2.1              \\ 
\end{tabular}
\caption{The results of our goodness-of-fit testing of our silica cloud polymorph forward models compared to the combined Hubble, Spitzer, and JWST data presented in \citet{Grant2023clouds}. Except where denoted ``only'', all models contain an additional deep pressure \ce{Al2O3} cloud deck. **\textit{We stress that these values should be used for demonstration only, as the optical properties used involve heavy extrapolations. For example, this poorer tridymite fit results from the most extrapolated wavelength region; the fit improves to 1.317 ($\chi^2/n$; $\chi^2$ = 125.2, BIC = 129.7) if 7.3 -- 8.1 $\mu$m is excluded.}}
\label{tab:stats}
\end{table}

\begin{table}[h]
\centering
\begin{tabular}{@{}lllllll@{}}
\toprule
Model                                         & $\chi^2$  & $\chi^2/n$ & BIC & $\Delta$BIC  \\ \midrule
\textit{Mie coefficients, BIC k = 1, n = 28, dof = 27}                     &      &     &                   \\
quartz                                        & 28.6  &     1.02 & 31.9    &  --                 \\
amorphous silica                              &  30.5 &   1.09 & 33.8    &   1.9                \\
tridymite                                     & 42.1 &    1.50** &  45.4   &   13.5            \\
cristobalite                                  &  31.5 &    1.13 & 34.9    &   3.0              \\ \midrule
\textit{Mie coefficients and density, BIC k = 2, n = 28, dof = 26}        &      &     &                   \\
quartz                                        & 28.6 & 1.02    & 35.3  &   3.4               \\
amorphous silica                              & 32.1& 1.14    & 38.7  &    6.8            \\
tridymite                                     &  45.9 & 1.64    &  52.5  & 20.6              \\
cristobalite                                  &  33.2 & 1.19   &  39.9  &  8.0                    \\ 
\end{tabular}
\caption{The results of our goodness-of-fit testing of our silica cloud polymorph forward models compared to only the JWST/MIRI LRS data presented in \citet{Grant2023clouds}. All models contain an additional deep pressure \ce{Al2O3} cloud deck. We do not include the additional $f_{\rm{sed}}$ parameter space fits from Table \ref{tab:stats}, as these are driven entirely by the optical slope and thus the MIRI data alone is not constraining, as detailed in \citet{Grant2023clouds}. **\textit{We stress that these values should be used for demonstration only, as the optical properties used involve heavy extrapolations. For example, the poorer tridymite fit results from the most extrapolated wavelength region; the fit improves to 1.39 ($\chi^2/n$; $\chi^2$ = 39.0) if 7.3 -- 8.1 $\mu$m is excluded.}}
\label{tab:stats_miri_only}
\end{table}

\newpage

\section{Optical Property Choices for Tridymite}

\begin{figure*}[h!]
\centering
{\includegraphics[width=.8\textwidth]%
{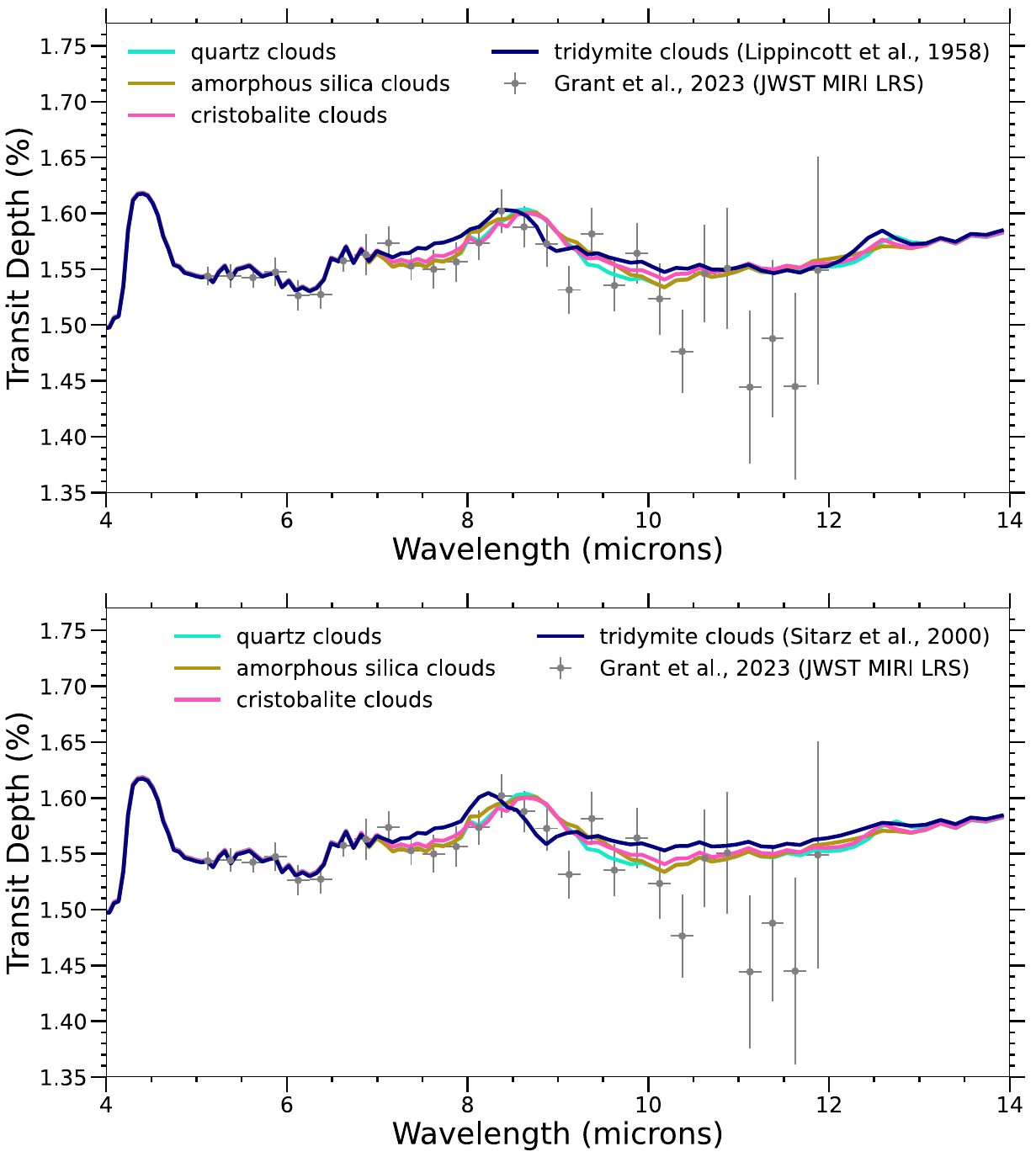}}
\caption{Additional polymorph models for WASP-17 b. We show here that the choice of whether to use the optical properties derived from \citet{lippincott1958infrared} or \citet{Sitzars2000_sio2_ir_spectra} impacts the observed \ce{SiO2} feature. All models use the best-fit values for the P-T profile, chemistry, and cloud parameters as in \citetalias{Grant2023clouds}. The main text uses only the \citet{Sitzars2000_sio2_ir_spectra} values, as these measurements were taken at elevated temperature. These differences highlight the need for careful, precise laboratoy measurements of this polymorph.}
\label{fig:tridymite_optprops}
\end{figure*}

\newpage
\section{Extended Model Runs for WASP-17 b}

\begin{figure*}[h]
\centering
{\includegraphics[width=0.92\textwidth]{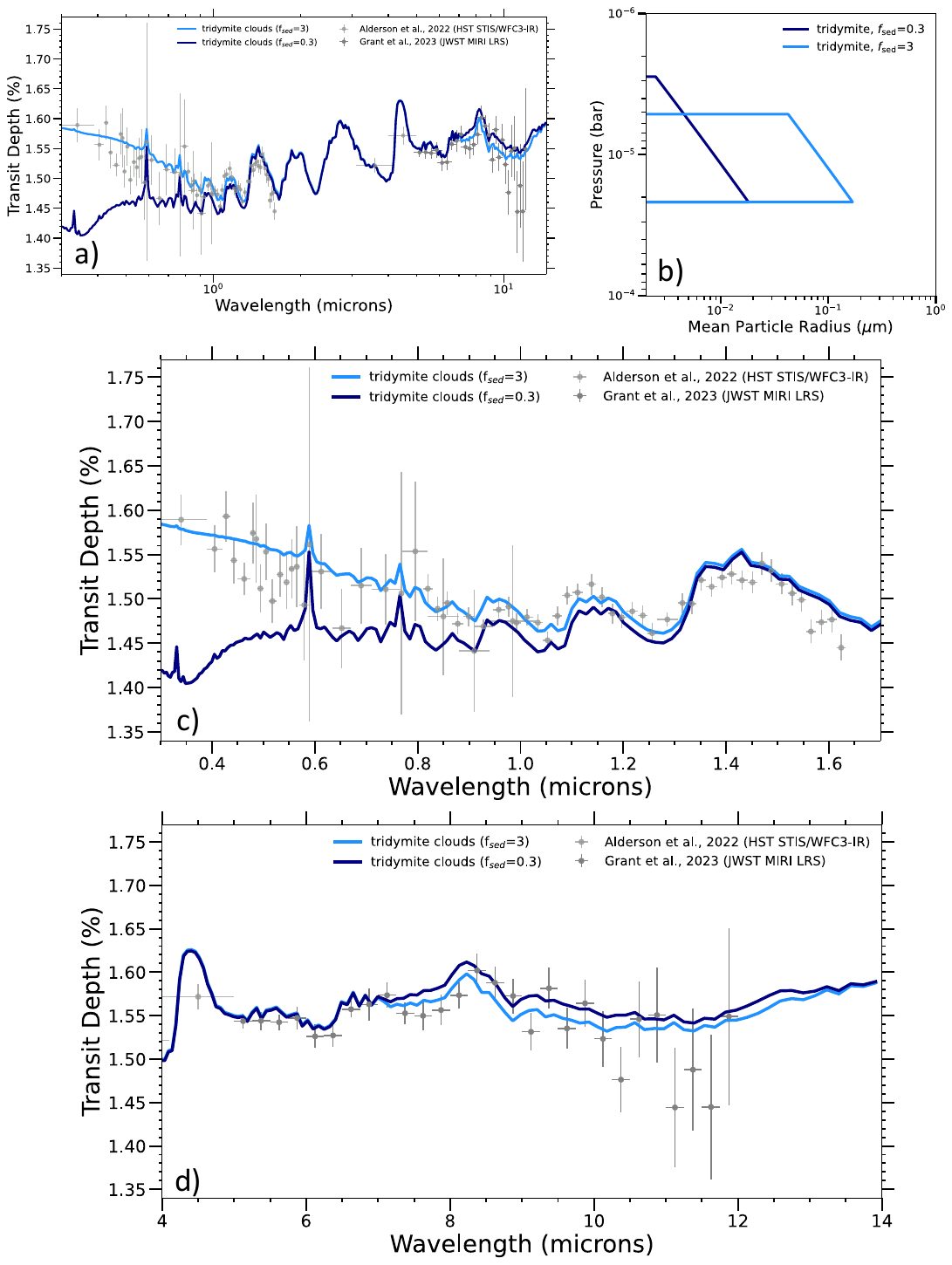}}
\caption{a) {\virga} atmospheric models of WASP-17 b using tridymite Mie coefficients and differing sedimentation efficiencies ($f_{\rm{sed}}$), with no alumina cloud layer. Dark blue lines show $f_{\rm{sed}}$=0.3, consistent with the best-fit {\virga} models of \citet{Grant2023clouds}; light blue shows $f_{\rm{sed}}$=3. b) Particle size distributions for the two different $f_{\rm{sed}}$ values, where $f_{\rm{sed}}$=0.3 is dark blue and $f_{\rm{sed}}$=3 is light blue. c) The same as (a), but focused on Hubble wavelengths. d) The same as (a), but focused on JWST/MIRI LRS wavelengths. 
\textbf{With different sedimentation efficiency, silica clouds alone can explain observations of WASP-17 b in {\virga} models without the need for a lower \ce{Al2O3} cloud deck.}}
\label{fig:fsed}
\end{figure*}

\end{document}